\documentclass[sigconf]{acmart}

\usepackage{epsfig}
\usepackage{balance}
\usepackage{amssymb}
\usepackage{amsmath}
\usepackage{amsfonts}
\usepackage{amsthm}
\usepackage{array,longtable}
\usepackage{graphicx}
\usepackage{booktabs}
\usepackage{subfig}
\usepackage{url}
\usepackage{color}
\usepackage{colortbl}
\usepackage{multirow}
\usepackage{caption}
\usepackage{fancyhdr}
\usepackage{enumerate}
\usepackage{xspace}

\copyrightyear{2019}
\acmYear{2019} 
\setcopyright{iw3c2w3}
\acmConference[WWW '19]{Proceedings of the 2019 World Wide Web Conference}{May 13--17, 2019}{San Francisco, CA, USA}
\acmBooktitle{Proceedings of the 2019 World Wide Web Conference (WWW '19), May 13--17, 2019, San Francisco, CA, USA}
\acmPrice{}
\acmDOI{10.1145/3308558.3313521}
\acmISBN{978-1-4503-6674-8/19/05}

\usepackage{soul}

\newcommand{\one}{({\em i}\/)}
\newcommand{\two}{({\em ii}\/)}

\def\eg{\emph{e.g.,}\xspace}

\def\ie{\emph{i.e.,}\xspace}
\def\etal{\emph{et al.}\xspace}
\def\vs{\emph{vs.}\xspace}

\newcolumntype{L}[1]{>{\raggedright\let\newline\\\arraybackslash\hspace{0pt}}m{#1}}
\newcolumntype{C}[1]{>{\centering\let\newline\\\arraybackslash\hspace{0pt}}m{#1}}
\newcolumntype{R}[1]{>{\raggedleft\let\newline\\\arraybackslash\hspace{0pt}}m{#1}}

\usepackage{amssymb}
\usepackage{pifont}

\usepackage{stackengine}
\setstackgap{L}{.5\baselineskip}

\usepackage{ifthen}

\begin{document}

\title{The Chain of Implicit Trust: An Analysis of the Web Third-party Resources Loading}
\author{Muhammad Ikram}
\email{muhammad.ikram@mq.edu.au}
\affiliation{%
  \institution{Macquarie University\\University of Michigan}
}

\author{Rahat Masood}
\email{rahat.masood@data61.csiro.au}
\affiliation{%
  \institution{UNSW and Data61, CSIRO}
}
\author{Gareth Tyson}
\email{g.tyson@qmul.ac.uk}
\affiliation{%
  \institution{Queen Mary University of London}
}
\author{Mohamed Ali Kaafar}
\email{dali.kaafar@mq.edu.au}
\affiliation{%
  \institution{Macquarie University and Data61, CSIRO}
}
\author{Noha Loizon}
\email{noha.loizon@data61.csiro.au}
\affiliation{%
  \institution{Data61, CSIRO}
}
\author{Roya Ensafi}
\email{ensafi@umich.edu}
\affiliation{%
  \institution{University of Michigan}
}
\renewcommand{\shortauthors}{Ikram et al.}
\sloppy

\begin{abstract}

The Web is a tangled mass of interconnected services, where websites import a range of external resources from various third-party domains. However, the latter can further load resources hosted on other domains. For each website, this creates a dependency chain underpinned by a form of implicit trust between the first-party and transitively connected third-parties. The chain can only be loosely controlled as first-party websites often have little, if any, visibility of where these resources are loaded from. 
This paper performs a large-scale study of dependency chains in the Web, to find that around 50\% of first-party websites render content that they did not directly load.
Although the majority (84.91\%) of websites have short dependency chains (below 3 levels), we find websites with dependency chains exceeding 30.
Using VirusTotal, we show that 1.2\% of these third-parties are classified as suspicious --- although seemingly small, this limited set of suspicious third-parties have remarkable reach into the wider ecosystem.
By running sandboxed experiments, we observe a range of activities with the majority of suspicious JavaScript downloading malware; worryingly, we find this propensity is greater among implicitly trusted JavaScripts. 

\end{abstract}
\maketitle

\section{Introduction}

In the modern web ecosystem, websites often load resources from a range of third-party domains such as ad providers, tracking services, content distribution networks (CDNs) and analytics services. 
This is a well known design decision that establishes an \textit{explicit trust} between websites and the domains providing such services. However, often overlooked is the fact that these third-parties can further load resources from other domains, creating a \emph{dependency chain}. This results in a form of \textit{implicit trust} between first-party websites and any domains loaded further down the chain.

Consider the \texttt{bbc.com} webpage, which loads JavaScript from the \texttt{widgets.com} domain, which, upon execution loads additional content from another third-party, say \texttt{ads.com}. Here, \texttt{bbc.com} as the first-party website, \emph{explicitly} trusts \texttt{widgets.com}, but \emph{implicitly}  trusts \texttt{ads.com}. This can be represented as a simple dependency chain in which \texttt{widgets.com} is at level 1 and \texttt{ads.com} is at level 2 (see Figure~\ref{fig:wot_repr}). Past work tends to ignore this, instead, collapsing these levels into a single set of third-parties~\cite{falahrastegar2014anatomy,Nikiforakis2012}. 
Here, we argue that this overlooks a vital aspect of website design. 
For example, it raises a notable security challenge, as first-party websites lack visibility on the resources loaded further down their domain's dependency chain. The dynamic nature of the content being loaded and the wide adoption of in-path traffic alterations~\cite{reis2008detecting,chen2016forwarding} further complicates the issue. This potential threat should not be underestimated as errant active content (\eg JavaScript) opens the way to a range of further exploits, \eg Layer-7 DDoS attacks \cite{Pellegrino2015} or massive ransomware campaigns \cite{Sequa2016}.

This paper studies dependency chains in the web ecosystem. Although there has been extensive work looking at the presence of third-parties in general~\cite{falahrastegar2014anatomy,Nikiforakis2012,Lauinger2017}, little work has focused on how content is indirectly loaded by first-party websites via dependency chains.
We start by inspecting how extensive dependency chains are across the Alexa's top-200K (Section \ref{sec:dataset})~\cite{alexa}. We confirm their prominence, finding that around 50\% of websites \emph{do} allow third-parties to form dependency chains (\ie they implicitly trust third-parties they do not directly load). The most commonly \emph{implicitly} trusted third-parties are well known operators, \eg  \texttt{google-analytics.com} and \texttt{doubleclick.net}: these are implicitly imported by 68.3\% (134,510) and 46.4\% (91,380) websites respectively. However, we also observe a wide range of more obtuse third-parties such as {\tt pippio.com} and {\tt 51.la} imported by 0.52\% (1,146) and 0.51\% (1,009) of websites. 
Although the majority (84.91\%) of websites have short chains (with levels of dependencies below 3), we find first-party websites with dependency chains exceeding 30 in length. This not only complicates page rendering, but also creates notable attack surface. 

With the above in mind, we then proceed to inspect if \emph{suspicious} or even potentially \emph{malicious} third-parties are loaded via these long dependency chains (Section \ref{sec:maldependency}).
We do not limit this to just traditional malware, but also include third-parties that are known to mishandle user data and risk privacy leaks. 
Using the VirusTotal service~\cite{VirusTotal} API, we classify third-party domains into innocuous \vs suspicious. When using a reasonable classification threshold, we find that 1.2\% of third-parties are classified as suspicious. 
Although seemingly small, we find that this limited set of suspicious third-parties have remarkable reach. 73\% of websites under-study load resources from suspicious third-parties, and 24.8\% of first-party webpages contain at least 3 third-parties classified as suspicious in their dependency chain. This, of course, is impacted by many considerations which we explore --- most notably, the power-law distribution of third-party popularity, which sees a few major players on a large fraction of websites. 

The prevalence of JavaScript resources being classified as suspicious leads us to further explore their activities. We therefore proceed to \textit{sandbox} all suspicious JavaScript programs to monitor their activities (Section \ref{sec:banalysis}). We build a sandbox and perform tests executing suspicious JavaScript. We observe that the further down the dependency chain, the more active the suspicious JavaScript are with a high number of HTTP requests generated by suspicious JavaScript programs at level $\geq$2. This is worrying as resources loaded further down the dependency chain are the most opaque to the website operator. We find evidence of involvement of first-party websites in malicious SEO (search) poisoning activities when (implicitly) loading some suspicious JavaScript content. Perhaps more importantly, we find that the most typical purpose of the suspicious JavaScript code is downloading dropfiles: again, the propensity to download files actually increases further along the chain with the most active JavaScript at level 4 downloading 129 files. %
We share our datasets, experimental testbed code and scripts used in this paper with the wider research community for further analysis of the consequences of the Web's implicit trust on {\tt https://wot19submission.github.io}. 
We conclude the paper by summarising reality of a very fragile web ecosystem, revealing that suspicious parties within the dependency chains are relatively commonplace (\S\ref{sec:conclusion}).

\section{Dataset and Data Enrichment}
\label{sec:dataset}

\subsection{Alexa dependency dataset}

We start by presenting our data collection methodology, and how we have validated its correctness. 

\subsubsection{Data Collection}
\label{subsec:static_datase}
\begin{figure}
\centering
{\includegraphics[width=0.48\textwidth]{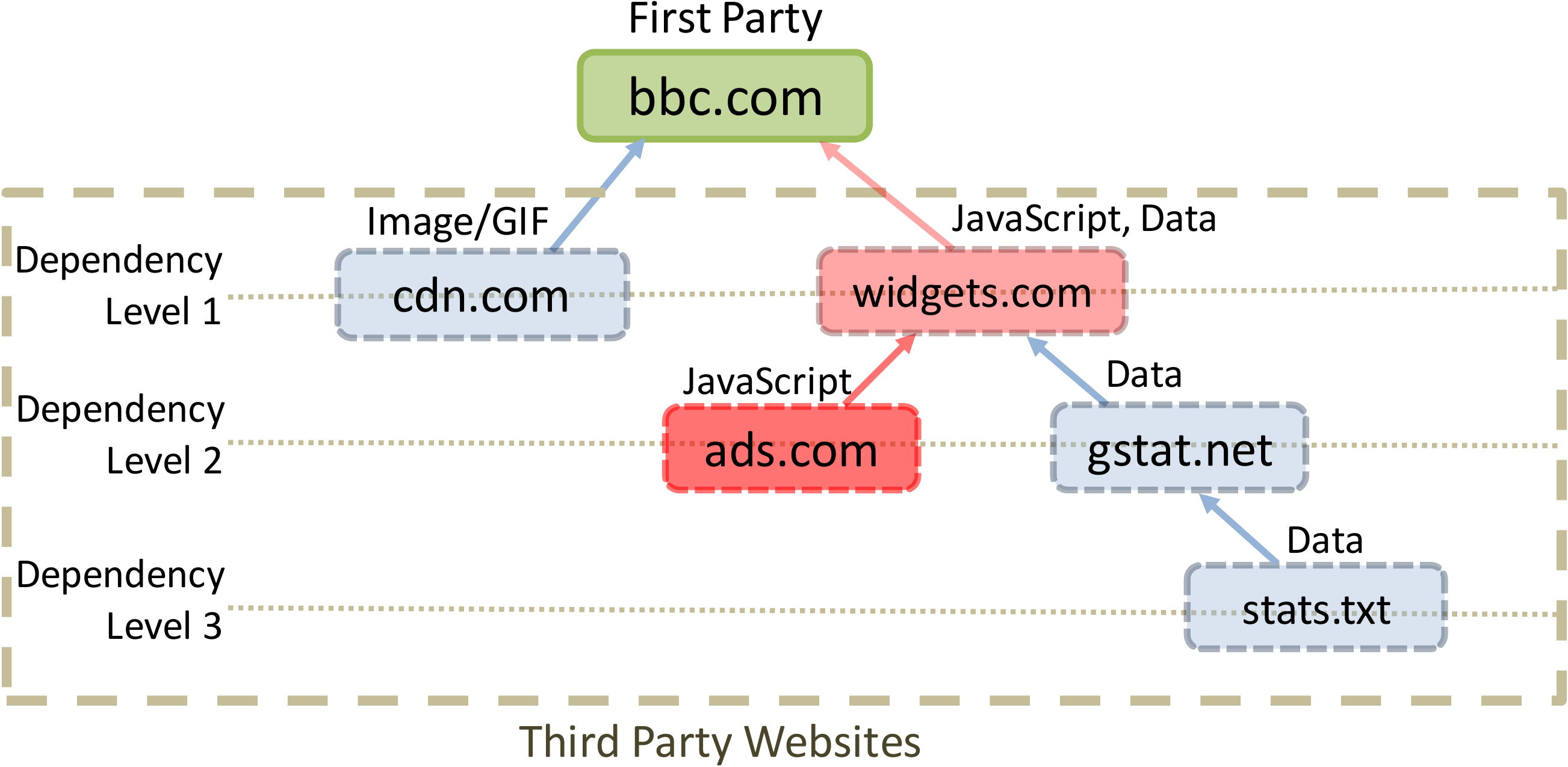}

\caption{Example dependency chain, including malicious third-party (in red). }%

\label{fig:wot_repr}}
\end{figure}

We obtain the resource dependencies of the Alexa top-200K websites' main pages\footnote{We select the top 200K as this gives us broad coverage of globally popular websites, whilst also remaining tractable for our subsequent data enrichment activities.} using the method described in~\cite{Kumar2017}. 
This Chromium-based Headless~\cite{gHeadless} crawler renders a given website and tracks resource dependencies by recording network requests sent to third-party domains. The requests are then used to reconstruct the dependency chains between each first-party website and its third-party URLs. Note that each first-party can trigger the creation of multiple dependency chains (to form a tree structure). 
Figure~\ref{fig:wot_repr} presents an example of a dependency chain with 3 levels; level 1 is explicitly trusted by the first-party website, whilst level 2 and 3 are implicitly (or indirectly) trusted. For simplicity, we refer to any domain that differs from the first-party to be a third-party. 
More formally, to construct the dependency tree, we identify third-party requests by comparing the second level domain of the page (\eg~\texttt{bbc.com}) to the domains of the requests (\eg~ \texttt{cdn.com} and \texttt{ads.com} via \texttt{widgets.com}). Those with different second level domains are considered third-party. We ignore the sub-domains so that a request to a domain such as {\tt player.bbc.com} is not considered as third-party. Due to the lack of purely automated mechanism to disambiguate between site-specific sub-domains (\eg~ {\tt player.bbc.com}) or country-specific sub-domains (\eg~ {\tt bbc.co.uk}), we leverage Mozilla Public Suffix list~\cite{mozslist} and {\tt tldextract}~\cite{tldextract} for this task.
From the Alexa Top-200k websites, we collect 11,287,230 URLs which consist of 6,806,494 unique external resources that correspond to 68,828 and 196,940, respectively, unique second level domains of third- and first-parties.

We acknowledge that the constructing the dependencies between objects in webpages is non-trivial task. In cases like where third-party JavaScript gets loaded into a first-party context, and then makes an AJAX request, the HTTP(S) request appears to be from the first-party (i.e. the {\it referrer} will be the first-party). To overcome such cases and to preserve the information on relations between the nested resource dependencies, we allow the crawler to include the URL of the third-party from which the JavaScript was loaded by first-party. 

\subsubsection{Data Validation}
\label{subsec:1500dailycrawl}

As our main dataset relies on a single snapshot, we want to evaluate the stability of the resources loaded by websites to ensure that a single snapshot does not miss significant complexity within the ecosystem. 
Thus, we repeat the methodology from Section~\ref{subsec:static_datase} on a daily basis. Unfortunately, performing daily crawls for the Alexa top-200k websites was not possible. We therefore selected 1,500 domains as a seed for the crawler. This list consisted of the Alexa top-1K alongside 250 domains randomly selected from the Alexa rank ranging from 1K to 50K, and a final 250 domains randomly chosen from websites within the Alexa rank 50K--200K. 
This offers a broad sampling of the Alexa sites covered. 
In total, on daily a basis from September 15--October 2 2018, we have collected on average 225,035 unique URLs per daily snapshot which covers 5,423 unique second level domains from the 1,500 first-parties. 

\begin{figure}[ht!]
\centering
	\includegraphics[width=0.48\textwidth]{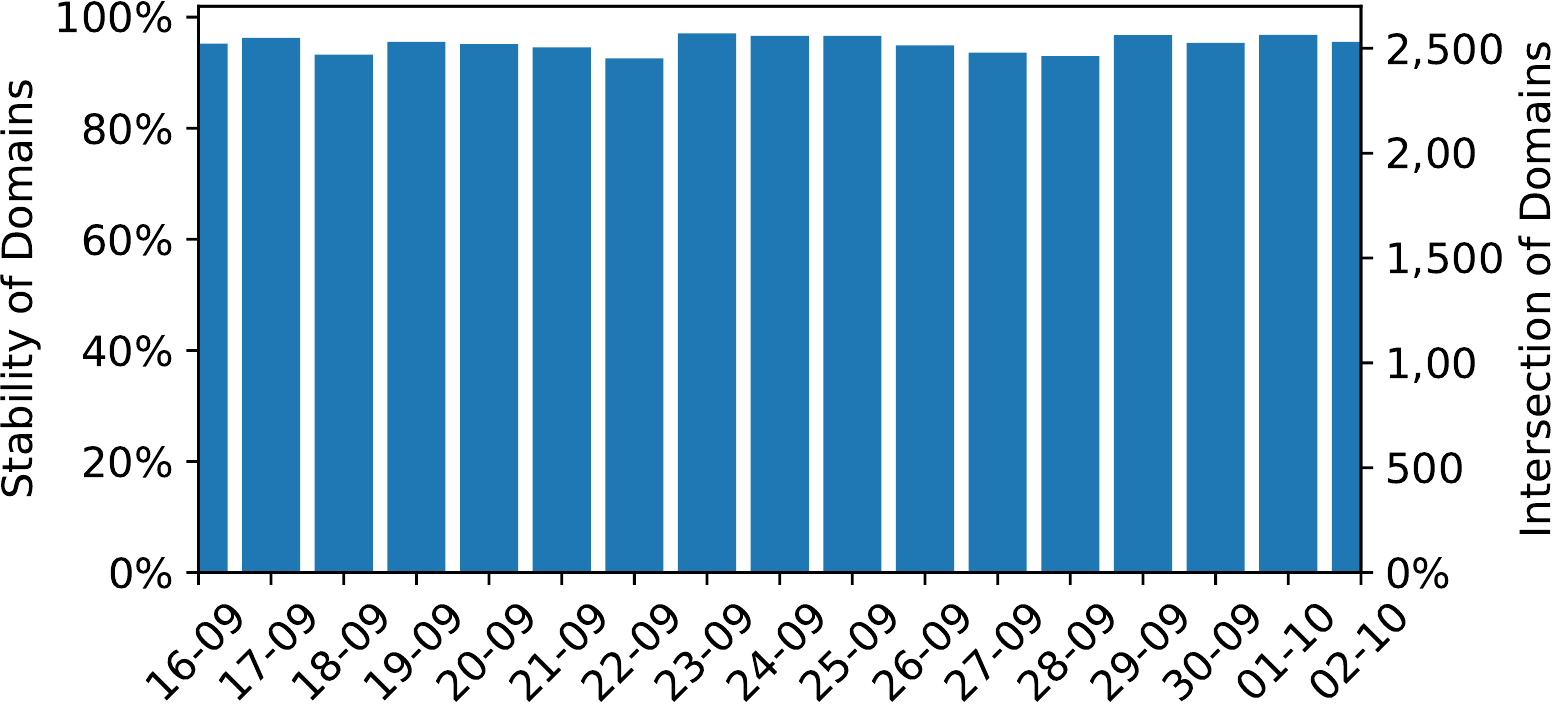}
\vspace{-0.5cm}
	\caption{\small Stability of day-by-day dependency trees analyzed per domain}
	\label{fig:intersection-n-stability}
\vspace{-0.6cm}	
\end{figure}

Figure~\ref{fig:intersection-n-stability} presents the day-to-day stability of the domains we see within each website.\footnote{We define the (normalized) stability as the count of domains present in the dependency trees crawled on day $n$ and also present on day $n~+~1$. More specifically, let $C$ denoting the crawled data then, stability $=$ $\frac{C_n \cap C_{n+1}}{C_n \cup C_{n+1}}$.} We observe that 95.07\% of second level domains remain consistent across consecutive days, and only an average 4.93\% of domains were absent in any two consecutive snapshots. 
On average, only 35 (0.66\%) and 232 (4.27\%) domains are absent at explicit and implicit dependency levels, respectively. Hence, we take this as strong indicator that utilizing a single snapshot is sufficient for gaining vantage into the use of third-parties. We leave inspection of temporal dynamics to our future work.

\subsection{Data Enrichment} 
\label{subsec:dataenrichment}

Malicious domains may circumvent a specific antivirus (AV) tool and, as suggested by previous studies~\cite{arp2014drebin, icse_2015-1}, some AV tools may not always report reliable results. Therefore, it is imperative to rely upon multiple AV scanners and datasets to effectively classify domains as innocuous \vs suspicious. We leverage the capabilities offered by VirusTotal's public API to automatize our classification process. VirusTotal is an online solution which aggregates the scanning capabilities provided by more than 68 AV tools, scanning engines and datasets. It has been commonly used in the academic literature to detect malicious apps, executables, software and domains~\cite{kantchelian2015better, Ikram:2016, kharraz2015cutting, Ibosiola19, ikram2017first}.

We use the VirusTotal \textsf{report} APIs to obtain the VTscore for each third-party URL. Concretely, this score is the number of AV tools that flagged the website as \textit{malicious} (max.\ 68). 
The reports also contain meta-information such as the first scan date, scan history, domain name resolution (DNS) history, website or domain category, reverse DNS, and whois information. 
We further supplement each domain with their WebSense~\cite{websense} category provided by the VirusTotal's \textsf{record} API. During the augmentation, we eliminate repeating, unresponsive or invalid URLs in each dependency chain. 
Thus, we collect the above metadata for each second level domain in our dataset.
This results in a final sample of 196,940 first-party websites, and 68,828 third-party domains.

\section{Exploring the Chains}
\label{sec:dependency}

We begin by exploring the presence and usage of implicit trust chains. We first confirm if websites do, indeed, rely on implicit trust and then explore how these chains are used. 

\subsection{Do websites rely on implicit trust?}

\begin{table}[ht!]
\centering
\small
\tabcolsep=0.05cm
\scalebox{0.95} {
\begin{tabular}{@{} l *9c @{}}\toprule&&&\multicolumn{1}{c}{\bf Alexa Rank}\\\cline{2-7}\multicolumn{1}{c}{} & 1-200K & 1-10K  & 190-200K & 10-50K  & 50-100K & 100-200K \\
\midrule
\small \textbf{F.-Parties that trust:} \\
\quad \small \textbf{All Resources:} \\
   \quad\quad  \small Explicit (Lvl. 1)    & 95\% &  95\% &  95\% & 94\% & 95\% & 95\%  \\
    \quad\quad   \small Implicit (Lvl. $\geq2$)    & 49.7\% &  55.1\% & 47.9\%  & 51.8\% & 50.23\% & 48\%  \\
    
\quad  \small \textbf{JavaScript:} \\
    \quad\quad   \small Explicit    & 91\% &  92\%   & 91\% & 91\% & 91\% & 90\%  \\
    \quad\quad   \small Implicit    & 49.5\% &  55\%  & 47.8\%  & 51.69\% & 50\% & 47.8\%  \\
    \bottomrule
 \end{tabular}
 }
 \caption{\small Overview of the Dataset for different ranges of Alexa's ranking. The rows indicate the proportion of Alexa's Top-X websites that explicitly and implicitly trust at least one third-party \one~resource (of any type); and \two~JavaScript.}
 \vspace{-0.5cm}
 \label{tab:dataset_ov}
\end{table}

Overall, the Top-200k dataset collectively makes 11,287,230 calls to 6,806,494 unique external resources, with a median of 27 external resources per first-party website. 
To dissect this, Table~\ref{tab:dataset_ov} presents the percentage of webpages in each Alexa range that load explicitly and implicitly trusted third-parties. Confirming several prior studies~\cite{falahrastegar2014anatomy,Lauinger2017}, it shows that 95\% of websites import external resources, with 91\% importing externally hosted JavaScript.
This trend is already well known; more important is that observation that around 50\% of the websites \emph{do} rely on implicit trust chains, \ie they allow third-parties to load further third-parties on their behalf. The propensity to form dependency chains is marginally higher in more popular websites; for example, 55\% in the Alexa top 10K have dependency chains compared to 48\%  in the bottom 10K (\ie rank 190-200K). 
In other words, more popular websites tend to rely more on implicitly trusted third-parties.

\begin{figure}[t]
\centering
\subfloat[]{
\includegraphics[width=0.24\textwidth]{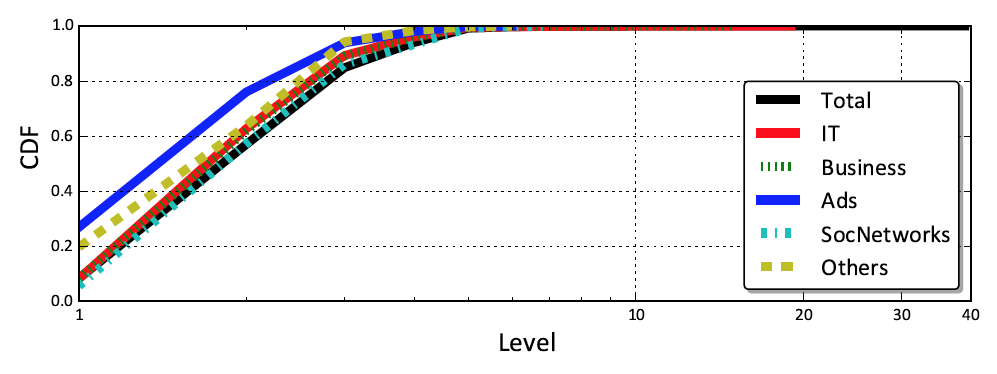}\label{fig:fp_levels}
}
\subfloat[]{
\includegraphics[width=0.24\textwidth]{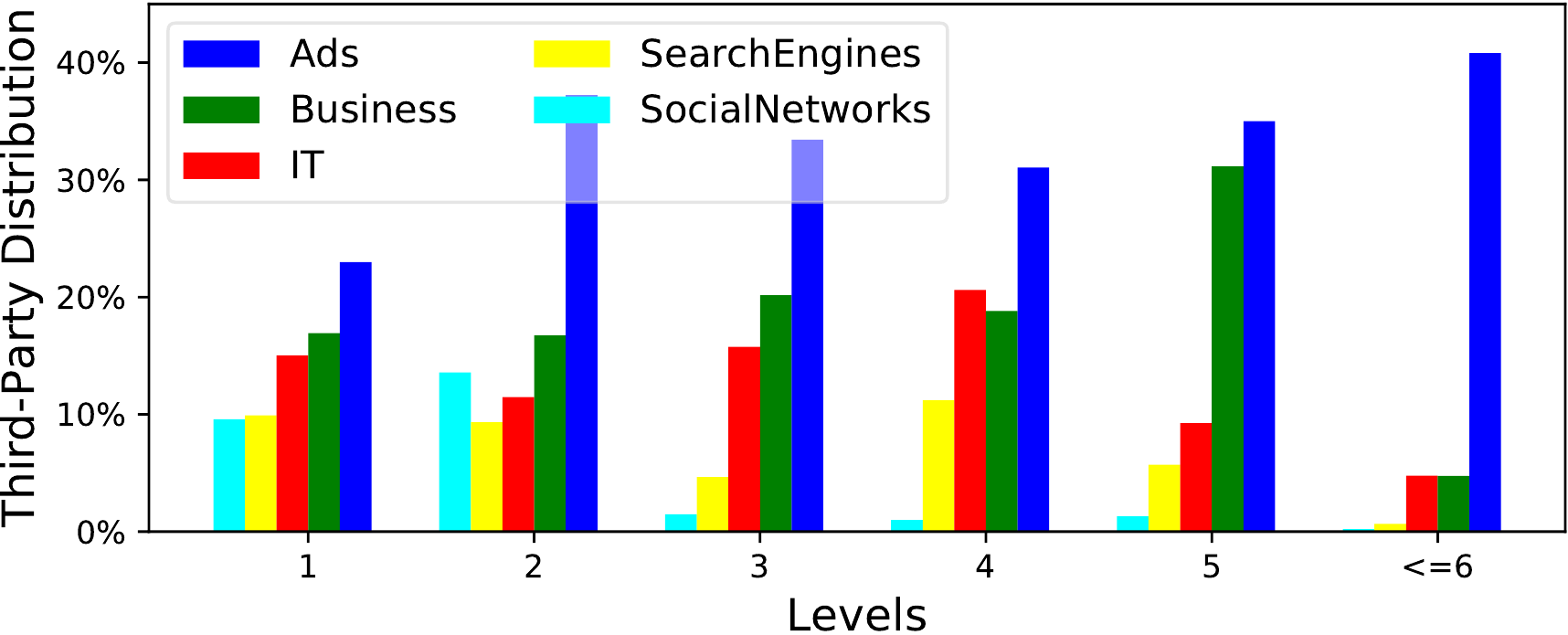}\label{fig:tp_levels}
}
\vspace{-0.2cm}
\caption{(a) CDF of dependency chain lengths (broken down into categories of first-party websites); and (b) distribution of third-party websites across various categories and levels.} 
\vspace{-0.3cm}
\end{figure}

These implicitly trusted third-parties appear at various positions in the dependency chain. 
Intuitively, long chains are undesirable as they typically have a deleterious impact on page load times~\cite{wang-nsdi13} and increase attacks surface.
Figure ~\ref{fig:fp_levels} presents the CDF of chain length for all first-party websites. For context, websites are separated into their sub-categories.\footnote{We only include the most popular categories.}
It shows that 80\% of the first-party websites create chains of trust of length 3 or below.  
However, there is also a small minority that dramatically exceed this chain length: we find that all website categories import $\approx$2\% of their external resources from level 3 and above.
In the most extreme case, we see \texttt{rg.ru} (news) with a chain containing 38 levels, consisting of mutual calls between \texttt{adriver.ru} (ad provider) and \texttt{admelon.ru} (IT website). Other notable examples include \texttt{thecrimson.com.bg} (Harvard's student newspaper), \texttt{argumenti.ru} (news), \texttt{mundomax.com} (IT news), \texttt{lifestyle.bg} (entertainment) have a maximum dependency level of 15. We argue that these complex configurations make it extremely difficult to reliably audit such websites, as a first-party cannot be assured of which objects are later loaded.

Briefly, we also note that Figure ~\ref{fig:fp_levels} reveals subtle differences \emph{between} different categories of third-party domains. For example, those classified as adverts are most likely to be loaded at level 1; this is perhaps to be expected, as many ad brokers naturally serve and manage their own content. In contrast, Social Network third-parties (\eg Facebook plug-ins) are least likely to be loaded at level 1. 

 \subsection{What objects exist in the chain?}
 
The previous section has confirmed that a notable fraction of websites create dependency chains with (up to) tens of levels. 
We next inspect the types of resources imported within these dependency chains. We classify resources into four main types: Image, JavaScript, Data (consisting of HTML, JSON, XML, plain text files), and CSS/Fonts. Table~\ref{tab:resource_types_layers} presents the volume of each resource type imported at each level in the trust chain. 
We observe that the make-up of resources varies dramatically based on the level in the dependency chain. For example, the fraction of images imported tend to increase --- this is largely because third-parties are in-turn loading images (\eg for adverts). 
In contrast, the fraction of JavaScript programs decreases as the level in the dependency chain increases: 30.6\% of resources at level 1 are JavaScript compared to just 12.3\% at level 3. This trend is caused by the fact that new levels are typically created by JavaScript execution (thus, by definition, the fraction of JavaScript must deplete along the chain). However, it remains at a level that should be of concern to web engineers as this confirms a significant fraction of JavaScript code is loaded from potentially unknown implicitly trusted domains.

\begin{table}[!t]
\small
\centering
\begin{tabular}{lcccccc}
\toprule
Lev. & Total  & Image & JS & Data & Font/CSS & Uncat. \\ 
\midrule
    1    & 9,212,245 &  34.4\% & 30.6\% & 16.0\% & 7.8\% &  11.3\% \\
    2    & 1,566,841 &  48.8\% & 16.7\% & 11.7\% & 3.3\% & 19.4\% \\
    3    & 405,390 & 45.0\% & 12.3\% & 11.1\% & 1.3\% & 30.2\% \\
    4    & 78,107 &  41.8\% & 18.4\% & 8.0\% & 8.1\% &  23.6\% \\
    5    & 14,413 &  40.6\% & 18.0\% & 12.8\% & 2.0\% & 26.4\% \\
    $\geq$6    & 10,208  & 36.6\% & 12.3\% & 13.0\% & 1.2\% & 36.8\% \\
    \bottomrule
 \end{tabular}
 \caption{\small Breakdown of resource types requested by the Top-200K websites across each level in the dependency chain. Total column refers to the number of resource calls made at each level. }
 \vspace{-0.8cm}
 \label{tab:resource_types_layers}
\end{table}

To build on this, we also inspect the \emph{categories} of third-party domains hosting these resources.
Figure~\ref{fig:tp_levels} presents the make-up of third-party categories at each level in the chain. It is clear that, across all levels, advertisement domains make up the bulk of third-parties. 
We also notice other highly demanded third-party categories such as search engines, Business and IT. 
These are led by well known providers, \eg \texttt{google-analytics.com} (web-analytics\footnote{Grouped as in business category as per VirusTotal reports.}) is on 68.3\% of pages. 
The figure also reveals that the distributions of categories vary across each dependency level. For example, 23.1\% of all loaded resources at level 1 come from advertisement domains, 37.3\% at level 2, 46.2\% at level 3, \ie the proportion increases across dependency levels. In contrast, social network third-parties (\eg Facebook) are mostly presented at level 1 (9.58\%) and 2 (13.57\%) with a significant drop at level 3. The dominance of advertisements is not, however, caused by a plethora of ad domains: there are far fewer ad domains than business or IT (see Table~\ref{tab:vt_resource_categories}). Instead, it is driven by the large number of requests to advertisements: Even though ad domains only make-up 1.5\% of third-parties, they generate 25\% of resources. Naturally, these are led by major providers. Importantly, these popular providers can trigger further dependencies; for example, \texttt{doubleclick.com} imports 16\% of its resources from further implicitly trusted third-party websites. This makes such domains an ideal propagator of malicious resources for any other domains having implicit trust in it.

 \section{Finding Suspicious Chains}
 \label{sec:maldependency}

The previous section has shown that the creation of dependency chains is widespread, and there is therefore extensive implicit trust within the web ecosystem. This, however, does not shed light on the activity of resources within the dependency chains, nor does it mean that the implicit trust is abused by third-parties. 
Thus, we next study the existence of \emph{suspicious} third-parties, which could lead to abuse of the implicit trust. Within this section we use the term \emph{suspicious} (to be more generic than malicious) because VirusTotal covers activities ranging from low-risk (\eg sharing private data over unencrypted channels) to high-risk (malware).

\subsection{Do chains contain suspicious parties?}
\label{subsec:third-partydomains}

\begin{table*}[ht!]
\centering
\small
\tabcolsep=0.08cm
\scalebox{1.0} {
\begin{tabular}{l c r r c  | c c  | c c  | c c | c c |c c}
\toprule
           &  & & & & \multicolumn{2}{c}{\bf VTScore $\geq$ 3} & \multicolumn{2}{c}{\bf VTScore $\geq$ 10} & \multicolumn{2}{c}{\bf VTScore $\geq$ 20}  & \multicolumn{2}{c}{\bf VTScore $\geq$ 40} & \multicolumn{2}{c}{\bf VTScore $\geq$ 55} \\
             \cline{5-15}
{\bf Category}  & {\bf  Third-Parties}     & {\bf Total Calls}  &{\bf  Suspicious JS}   &   & Num. &  Vol. & Num. &  Vol. & Num. &  Vol. & Num. &  Vol. & Num. &  Vol. \\

\midrule
All  & 68,828  & 11,287,204        & 270,758 (2.4\%)   &       &  1.6\%      & 6.4\%   & 1.2\% & 6.2\% & 1.0\%   & 6.1\%  & 0.6\%	 & 5.7\%	 & $\leq0.1\%$   & $\leq0.1\%$      \\
Business  & 6,786  & 1,924,591        & 184,360 (9.6\%)  &        &  1.5\%      & 	21.5\%   & 1.1\% & 21.5\% & 1.0\%          & 21.4\%   & 0.5\%          & 20.6\%     & 0\%   & 0\%      \\
Ads  & 1,017  &		2,870,482        & 7,924 (0.3\%) &    &  3.5\%      & 	0.1\%   & 3.3\%   & 0.1\%    & 2.9\% & 0.1\% & 1.6\%     & $\leq0.1\%$     & 0\%   & 0\%      \\
IT  & 8,619  & 1,646,287        & 10,547 (0.6\%) &     &  2.2\%      & 	3.8\%   & 1.5\% 	& 3.6\%		 & 1.2\%          & 3.5\%   & 0.6\%          & 3.0\%     & $\leq0.1\%$   & $\leq0.1\%$      \\
Other & 52,406  & 4,845,844        & 67,927 (1.4\%) &      &  1.4\%      & 	4.6\%   & 1.1	 & 4.3\%	 & 0.9\%          & 4.2\%   & 0.6\%          & 3.8\%     & $\leq0.1\%$   & $\leq0.1\%$      \\ 

\bottomrule
\end{tabular}
}
\caption{\small Overview of suspicious third-parties in each category. \textbf{Col.2-4:} number of third-party websites in different categories, the number of resource calls to resources, and the proportion of calls to suspicious JavaScript. \textbf{Col.5-9:} Fraction of third-party domains classified as suspicious (\textit{Num.}), and fraction of resource calls classified as suspicious (\textit{Vol.}), across various VTscores (i.e., $\geq$ 3 and $\geq$ 55). }
\label{tab:vt_resource_categories}
\end{table*}

First, we inspect the fraction of third-party domains that trigger a warning by VirusTotal. From our third-party domains, 2.5\% have a VTscore of 1 or above, \ie at least one virus checker classifies the domain as suspicious. If one treats the VTscore as a ground truth, this confirms that popular websites \emph{do} load content from suspicious third-parties via their chains of trust. However, we are reticent to rely on VTscore $\geq1$, as this indicates the remaining 67 virus checkers did not flag the domain\footnote{Diversity is likely caused by the virus databases used by the different virus checkers~\cite{canto2008large}}. Thus, we start by inspecting the presence of suspicious third-parties using a range of thresholds. 

Table~\ref{tab:vt_resource_categories} shows the fraction of third-parties that are classified as suspicious using several VTscore thresholds. For context, we separate third-parties into their respective categories (using WebSense). The table confirms that a noticeable subset of suspicious third-party domains exist; for example, if we classify any resource with a VTscore $\geq10$ as suspicious, we find that 1.2\% of third-party domains are classified as suspicious with 6.2\% of all resource calls in our dataset going to these third-parties. Notably this only drops marginally (to 5.7\%) with a \emph{very} conservative VTscore of $\geq40$. We observe similar results when considering thresholds in the $[3  .. 50]$ range. This confirms, with a high certainty, that approximately 6\% of resource calls in the dependency chains are towards domains that engage in suspicious activity (see Section~\ref{sec:banalysis}) for further details). We will conservatively refer to domains with a VTscore $\geq 10$ as suspicious in the rest of this analysis.

Additionally, we inspect first-party domains that inherit suspicious JavaScript resources from the explicit and various implicit levels. We focus on JavaScript as active web content that poses great threats with significant attack surfaces consisting of vulnerabilities related to client-side JavaScript, such as cross-site scripting (XSS) and advanced phishing \cite{Lauinger2017}. 
Table~\ref{tab:mal_firstparty2} shows top first-party domains, ranked according to the number of unique suspicious third-parties in their chain of dependency. We note that the top ranked (most vulnerable) first-party domains belong to various categories such as Content Sharing, News, or IT. This indicates that there is no one category of domains that inherits suspicious JavaScript. However, we note that first party websites categorized as ``Business'' represent the majority of most exposed domains at Level $\geq $2, with 16\% of the total number of first-party domains implicitly trusting suspicious JavaScript belonging to the Business Category, with distant second being the  ``News \& Media'' Category and third the ``Adult'' category. The number of suspicious JavaScript codes loaded by these first-party domains ranges from 4 to 31 programs. We note the extreme case of \texttt{amateur-fc2.com} website \textit{implicitly} importing 31 unique suspicious JavaScript programs from 4 unique suspicious domains. Moreover, we observe at most 7 unique third-parties (combining both explicit and implicit level) that is a cause of suspicious JavaScripts in first-parties.  This happens for \texttt{privet-rostov.ru} domain, having third-party domains such as \texttt{charter.com, vk.com. rambler.ru, doubleclick.net, dx.com, cdn.adlegend.com, syncsw.pool.datamind.ru}.

\begin{table}[ht!]
\centering
\small
\tabcolsep=0.08cm
\scalebox{0.9}{
\begin{tabular}{l l r c c l c}
\toprule
\multicolumn{7}{c}{\bf Unique Suspicious Domains at Level $=$ 1} \\
\midrule
&	&	{\bf Alexa}	&	{\bf $\#$ Mal.} 	&	{\bf Unique}	&	& {\bf Chain}\\
$\#$	&	{\bf First-party Domain}&	{\bf Rank}	&	{\bf JSes}	&	{\bf Susp. Doms.}	&	{\bf Category}& {\bf Len.}\\
\midrule
1	&	theinscribermag.com	&	46,242	&	6	&	5	&	Blogs & 5\\
2	&	skynet-system.com.ua	&	192,549	&	6	&	5	&		Busin. & 4\\
3 & nodwick.com  & 194,823 & 13  & 4 &  Enter.& 4\\ 
4  & iphones.ru  & 12,045 & 4  & 4 &   IT  & 4\\ 
5	&	privet-rostov.ru  & 193,024	&	6	&	4	&	LifeStyle & 4\\
\midrule
\multicolumn{6}{c}{\bf Unique Suspicious Domains at Level $\geq$ 2} \\
\midrule
1  & {traffic2bitcoin.com}   & {33,513}  & {6} &{5}  & Games & 7\\ 
2  & {radionetplus.ru} & {166,003}  & {8}  & {4}  & SW Download & 6\\
3  & {studiofow.tumblr.com}   & {85,483}  & {11} & {4}  & Adult & 4\\  
4  & {amateur-fc2.com}   & {52,556}  &{31} & {4}  & Adult & 5\\  
5 & {fasttorrent.ru} & {24,250} & {9}  & {4}  &  File Sharing & 7\\  

\bottomrule
\end{tabular}
}
\caption{\small Top 5 most exposed first-party domains (with VTscore $\geq$ 10) ranked by the number of unique suspicious domains. }
\label{tab:mal_firstparty2}
\end{table}

\subsection{How widespread are suspicious parties?}

\begin{figure}[t]
\centering
\subfloat[All websites]{\includegraphics[width=0.24\textwidth]{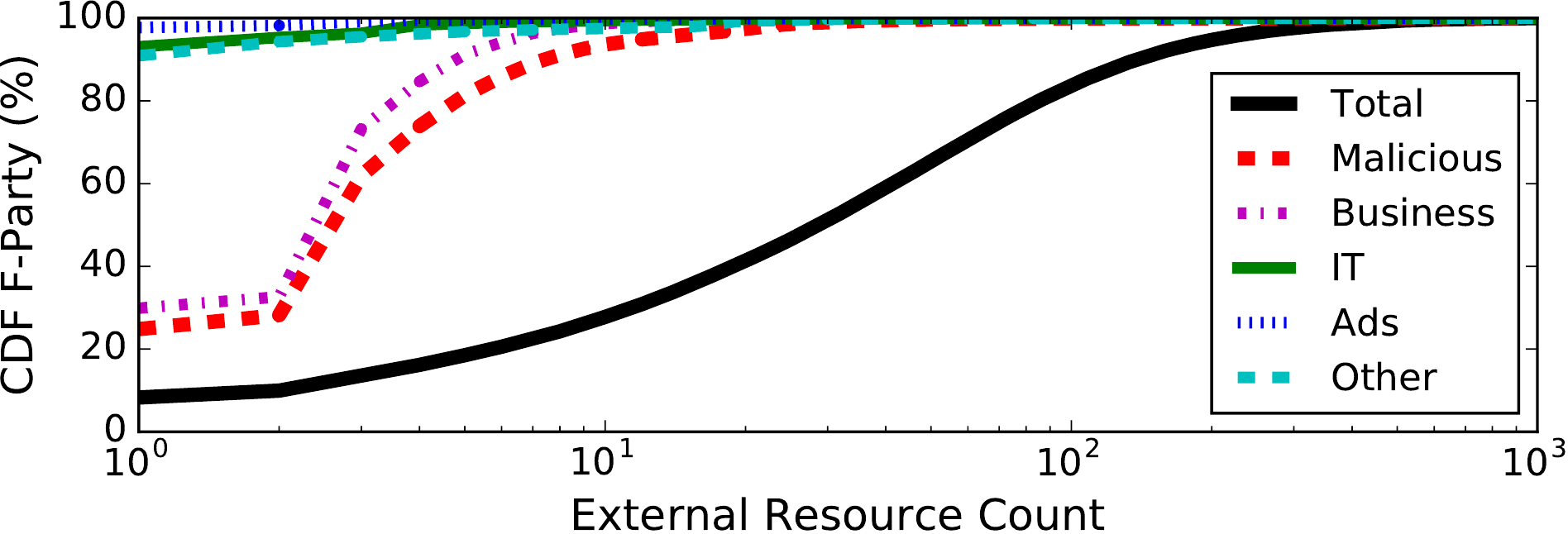}\label{fig:all_tpcat_cdf}}
\subfloat[Without google-analytics]{\includegraphics[width=0.24\textwidth]{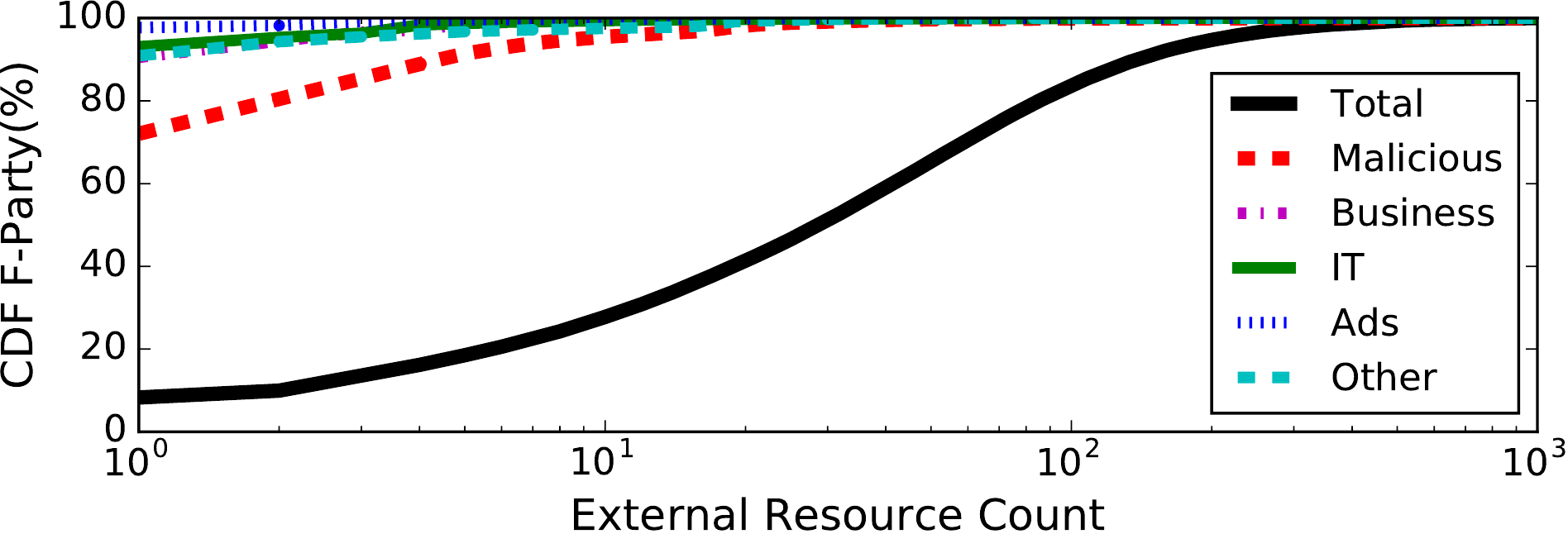}\label{fig:no_googlea_tpcat_cdf}}
\caption{\small CDF of resources loaded per-website from various categories of third-parties.}
\label{fig:mal_tpcat_cdf}
\end{figure}

We next inspect how ``popular'' these suspicious third-parties are at each position in the dependency chain, by inspecting how many websites utilize them. 
Figure~\ref{fig:all_tpcat_cdf} displays the cumulative distribution (CDF) of resource calls to third-parties made by each first-party webpage in our dataset. Within the figure, we decompose the third-party resources into various groups (including total \vs suspicious). As mentioned earlier, we take a conservative approach and consider a resource suspicious if it receives a VTscore $\geq$ 10.
The figure reveals that suspicious parties within the dependency chains are commonplace: 24.8\% of all first-party webpages contain at least 3 third-parties classified as suspicious in their dependency chain. Remarkably, 73\% of first-party websites load resources from third-parties at least once.
Hence, even though only 1.6\% of third-party domains are classified as suspicious, their reach covers nearly three quarters of websites (indirectly via implicit trust). 

This is a product of the power-law distribution of third-party ``popularity'' across websites: The top 20\% of third-party domains cover 86\% (9,650,582) of all resource calls. 
Closer inspection shows that it is driven by one prominent third-party: \texttt{google-analytics.com}. At first, we thought that this was an error, however, during the measurement period \texttt{google-analytics.com} obtained a VTscore of 51, suggesting a high degree of certainty. This was actually caused by \texttt{google-analytics.com} loading another third-party, \texttt{sf-helper.net}, which is known to distribute adwares and spywares. It is unclear why Google was performing this.
We therefore repeated these checks in October 2018, to confirm that this activity has ceased, and \texttt{sf-helper.net} is no longer loaded. Hence to understand the impact its new de-classification has, Figure~\ref{fig:no_googlea_tpcat_cdf} shows the distribution of resource calls to third-party categories when \texttt{google-analytics.com} is benign.
This reduces the number of first-party websites exposed to suspicious resources by 63\%. This highlights effectively the impact of high centrality third-parties being permitted to load further resources: the infection of just one can immediately effect a significant fraction of websites.

\begin{table}[ht!]
\centering
\small
\tabcolsep=0.08cm
\scalebox{0.95}{
\begin{tabular}{l l r c l}
\toprule
\multicolumn{5}{c}{\bf Prevalence of Third-parties at Level = 1} \\
\midrule
$\#$	&	{\bf \small Third-party Domain}&	{\bf Alexa Rank}	&	{\bf \# FP}	& {\bf Category}\\
\midrule
1	&	google-analytics.com	&	13,200	&		43,156	&		Business (Web Analytics)\\
2	&	gravater.com	&	2,292	&	3,520	&		IT\\
3	&	charter.com	&	12,714	&	3,425	&	Business\\
4	&	vk.com	&	13	&	2,815 	&		Social Network\\
5	&	statcounter.com	&	2,265	&	2,327		&	Business (Web Analytics)\\
\midrule
\multicolumn{5}{c}{\bf \small Prevalence of Third-parties at Level $\geq$ 2} \\
\midrule
1  & {charter.com}   & {12,714}  &{3,452} &Business\\  
2  & {vk.com}   &   {13} & {2,290} &  Social Network\\  
3  & {livechatinc.com}   & {888}  & {851}  & Web Chat\\  
4  & {onesignal.com}   & {950}  & {467} &  Business\\
5  & {rambler.ru}   & {291}  &  {370} & SearchEngine\\  

\bottomrule
\end{tabular}
}
\caption{\small Top 5 most prevalent suspicious third-party domains (with VTscore $\geq$ 10) on level 1 (explicit trust) and beyond (implicit trust) providing resources to first-parties. \#FP refers to the number of First-party domains having the corresponding suspicious third-party domain in their chain of dependency.}
\label{tab:top10_tp_lvl1}
\end{table}

Next we inspect in, Table~\ref{tab:top10_tp_lvl1}, the top 10 most frequently encountered suspicious third-party domains that are providing suspicious JavaScript resources to first-parties (as opposed to the most exposed first-party domains shown earlier in Table~\ref{tab:mal_firstparty2}). We rank these suspicious third-party domains according to their prevalence in the Web ecosystem and further decompose our analysis at explicit and implicit levels in the table.

As already discussed, we found {\tt google-analytics.com} among the most called domains. Interestingly, we find several suspicious third-party domains from the Top 100 Alexa ranking. 
For-instance, \texttt{vk.com}, a social network website mostly geared toward East-European countries has been used by 3,094 first-parties and is ranked 13 by Alexa. This website is found to be one of the most prevalent suspicious third-party domains at both level 1 and levels $\geq$ 2. An obvious reason for this domain's presence is because of other infected (malware-based) apps that try to authenticate users from such domains~\cite{vk_ref}. 
Other websites such as \texttt{statcounter.com} or \texttt{gravater.com} are also among the most prevalent third party domains in level 1. These websites were reported to contain malware in their Javascript codes~\cite{stat_ref}. For instance, users in statcounter forums reported it as malicious because a Javascript  running its website redirects users to a malware website \texttt{gocloudly.com}, and forces users to click the button ~\cite{stat_ref2}. 

While it is not shown in the tables, we also note the presence of {\tt qq.com}, a Chinese Search Engine ranked high by Alexa. This is among the top 10 most encountered suspicious third-party domains, as defined by several AntiVirus tools within VirusTotal. Closer inspection reveals this is likely due to repeated instances of insecure data transmission, use of qq fake accounts for malware manifestation and for data encryption Trojans~\cite{qq_ref, qq_ref2, qq_ref3}.

\begin{figure}[th!]
\centering
	\subfloat[]{\includegraphics[width=0.24\textwidth]{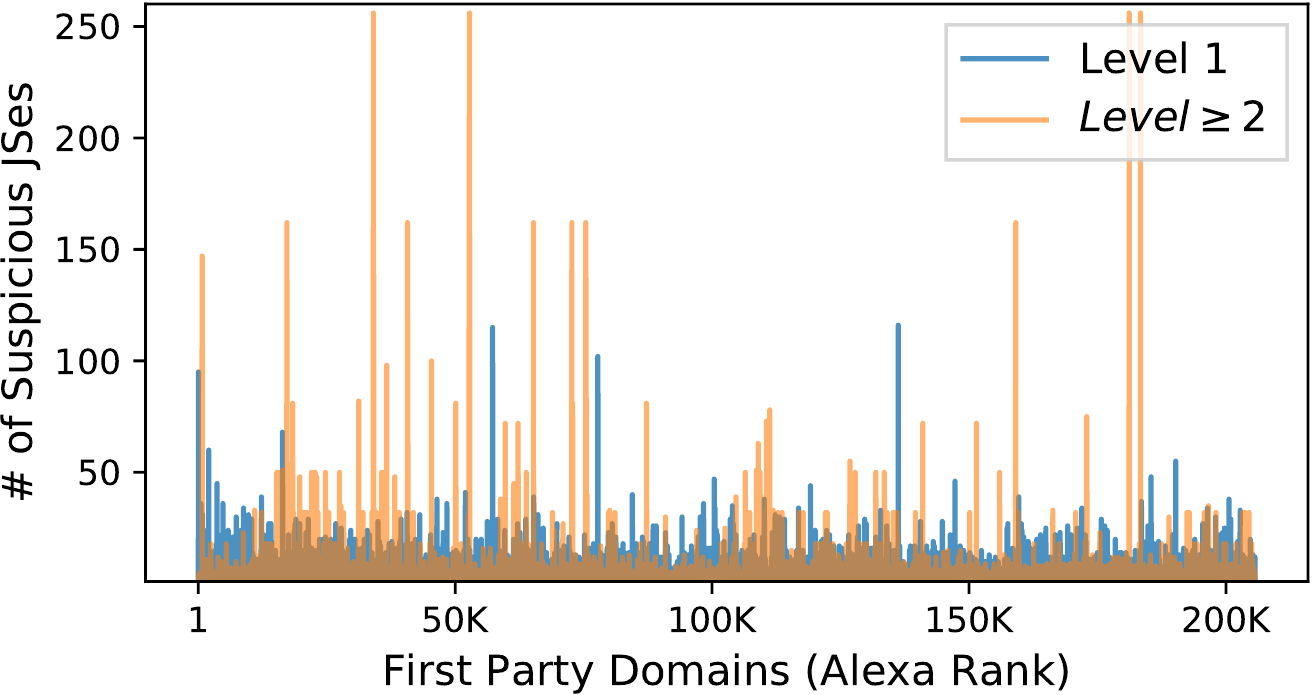}\label{Fig:FP-JSes}}
	\subfloat[]{\includegraphics[width=0.24\textwidth]{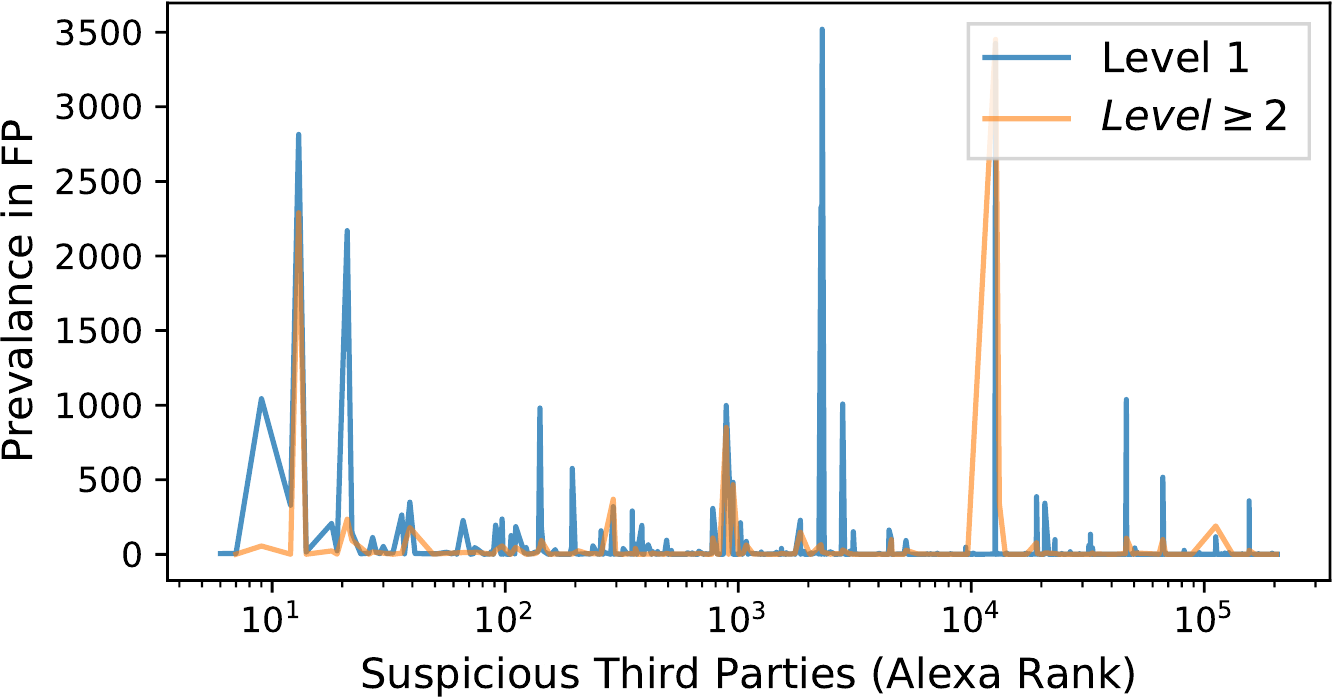}\label{Fig:TP-Prev}}
	\caption{\small Figure (a) depicts the number of suspicious JavaScript content imported (explicitly and implicitly) by first-party domains shown according to their Alexa ranking; and (b) shows the number of impacted first-party domains as function of the ranking of domains of Suspicious JavaScript. }
\end{figure}

More generally, we observe the presence of a wide range of Alexa ranks in the list of most prevalent domains at levels $\geq$ 2. 
In Figure ~\ref{Fig:FP-JSes}, we show the number of suspicious JavaScripts imported by the first-party domains (Y-axis) according to their Alexa rank (X-axis). Overall, first-party domains import a larger number of suspicious third-party JavaScript codes at levels $\geq$ 2, however, the first-party domains seem to be equally vulnerable to the implicit import of suspicious content regardless of their rank.
There are exceptions though, signified by the peaks in the number of suspicious JavaScripts --- these are near exclusively driven by a large number of $\geq$level-2 scripts (implicit trust). 
We also encountere an interesting case, which we exclude from the graphs for readability purposes: The first-party domain \texttt{kikar.co.il} imports 2,592 JavaScript codes originating from the third-party \texttt{hwcdn.net}, a well-known browser hijacker that has been reported to force users to visit spam pages~\cite{hwcdn_ref}. The VirusTotal API indicates a VTscore of 22 for this suspicious domain. We also note that 35 other first-party domains have this domain in their chain of dependency. Again, this highlights the risk of implicit trust. 

In Figure \ref{Fig:TP-Prev} we show the number of impacted first-party domains as a function of the Alexa Rank of suspicious third-party domains (limited to a maximum Alexa Rank of 1 million) --- note the log scale of x axis.  We observe that some very prevalent third-parties have a high Alexa ranking (even excluding \texttt{google-analytics.com} which as per our previous observation has the highest prevalence of 43,156 impacted first-party domains and hence excluded from Figure~\ref{Fig:TP-Prev} for readability purposes). Note a spike around 2000 rank, which reaches a prevalence of 3500 first-party domains at level 1. This spike is caused by \texttt{gravatar.com}, propagating suspicious Javascripts. This supports our statements earlier (from Table~\ref{tab:top10_tp_lvl1}) where \texttt{gravatar.com} is ranked second top most suspicious domain. Similarly, a spike around 10K rank indicates the presence of \texttt{charter.com} both at level 1 and 2 respectively. These findings demonstrate the wide variety of third-party suspicious JavaScript content loaded from various, not necessarily ``obscure'', third-party domains.

\subsection{Which websites are impacted?}

\begin{figure*}[ht!]
\centering
\subfloat[All websites]{\includegraphics[width=0.31\textwidth]{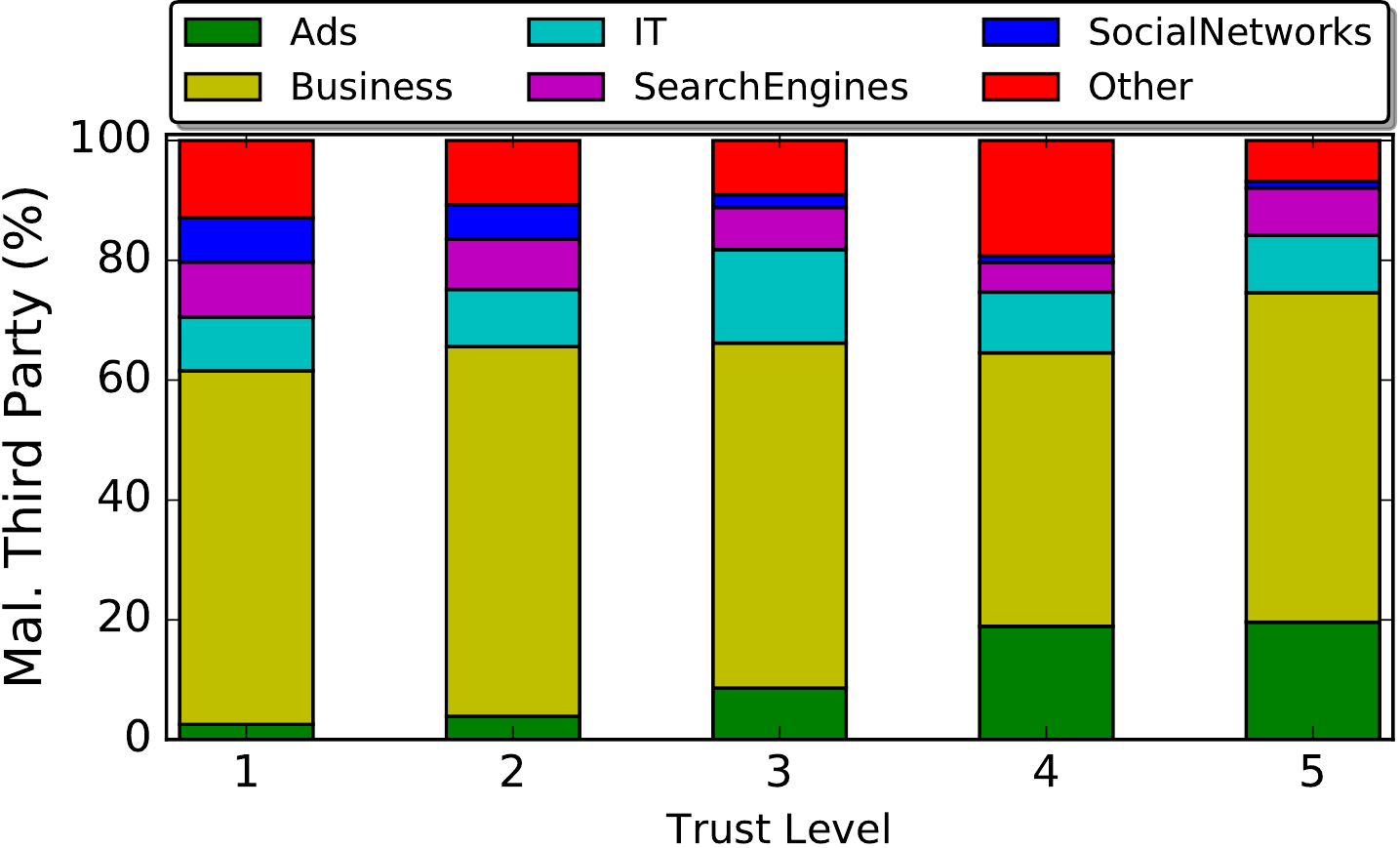}\label{fig:all_fp_stack}}
\quad
\subfloat[News Websites]{\includegraphics[width=0.31\textwidth]{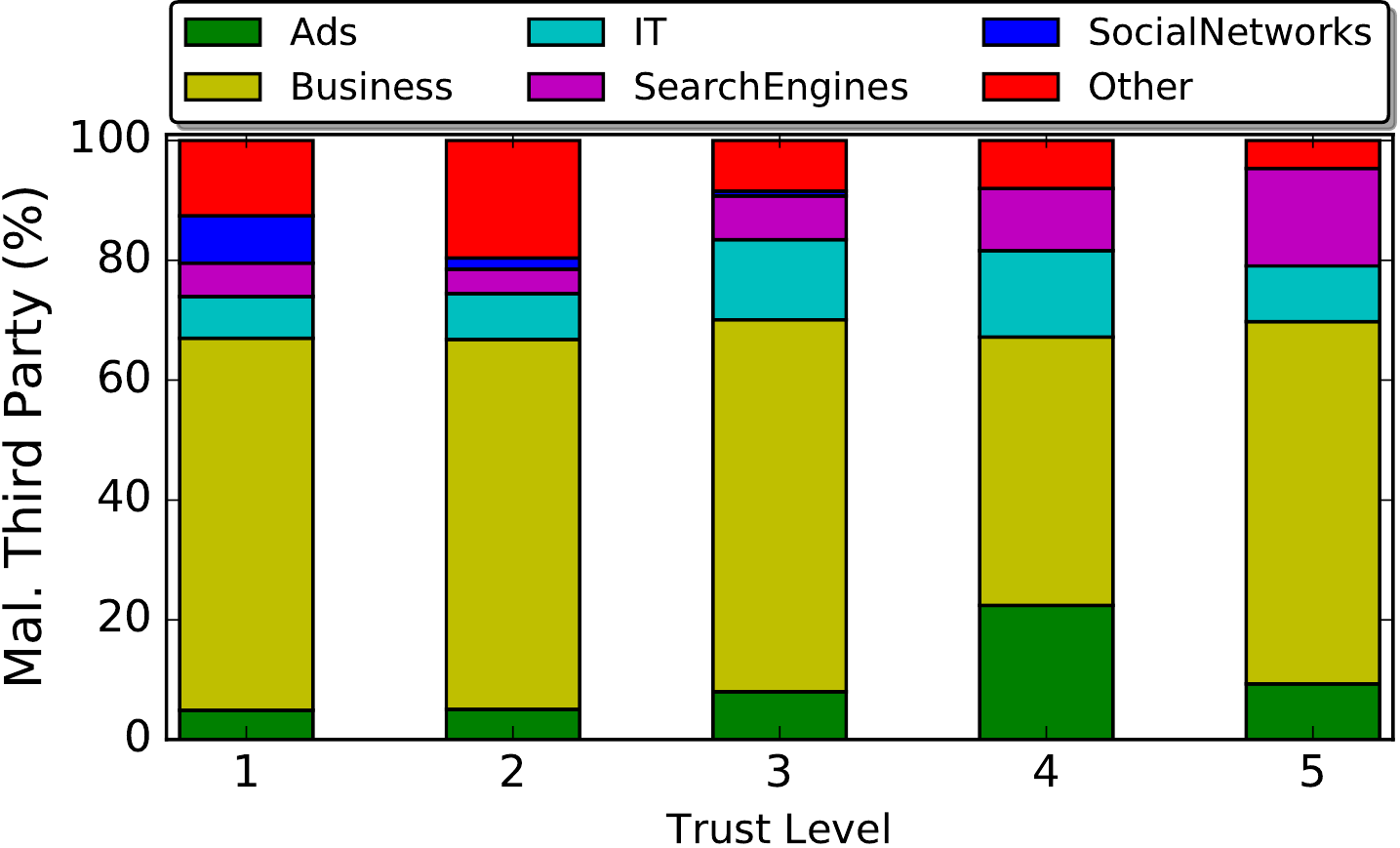}\label{fig:news_fp_stack}}
\quad
\subfloat[Sports websites]{\includegraphics[width=0.31\textwidth]{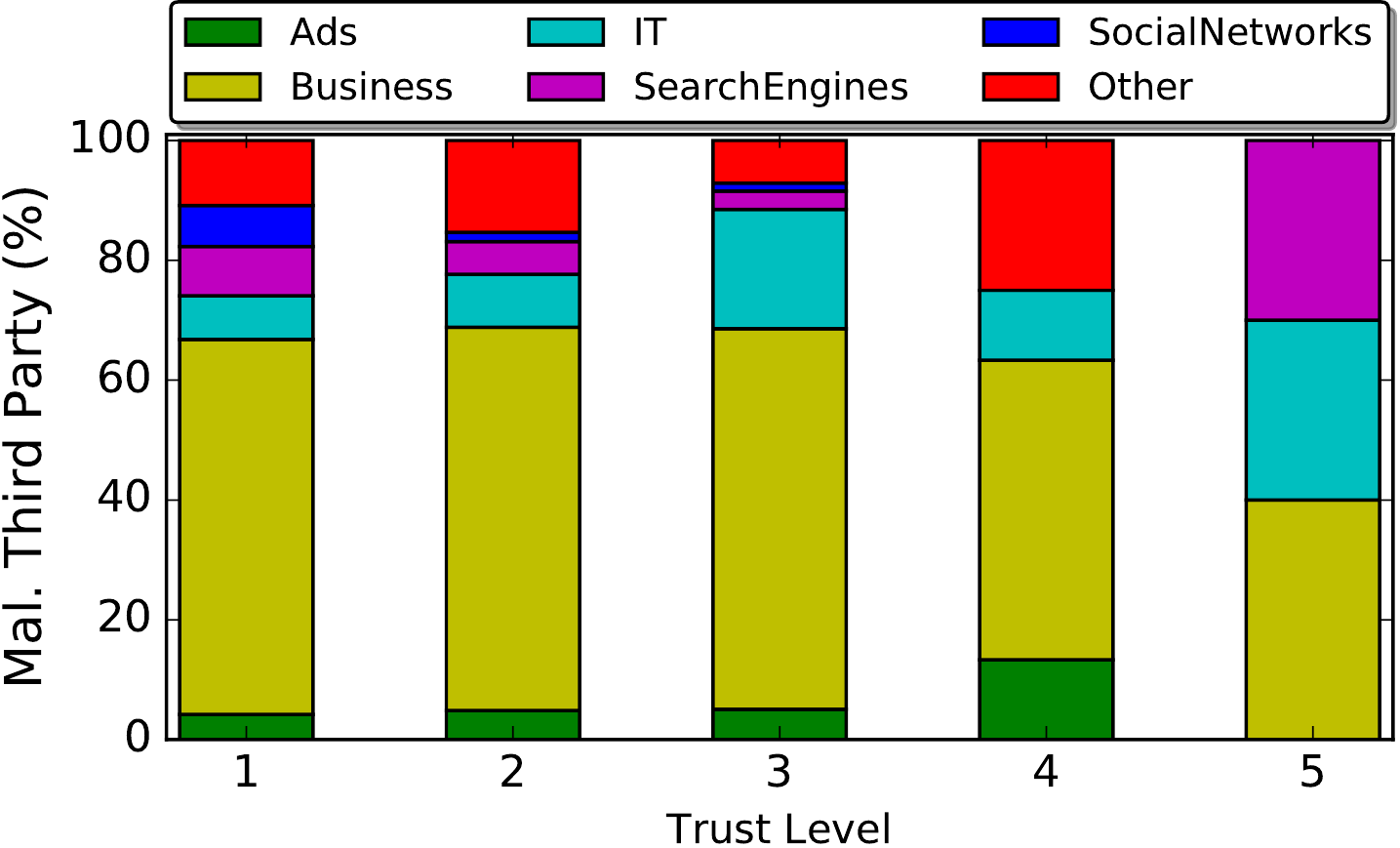}\label{fig:sports_fp_stack}}
\caption{ \small Distribution of calls to suspicious third-party websites per category at each level, for all top-200K websites (Figure~\ref{fig:all_fp_stack}) and most vulnerable first-party categories (Figures~\ref{fig:news_fp_stack},~\ref{fig:sports_fp_stack}). } 
\label{fig:fp_resource_bar}
\end{figure*}

Next, we inspect the location(s) in the dependency chain where these suspicious third-parties are situated, as well as the \emph{types} of websites that load them.
This is vital, as implicitly trusted ($\geq$level 2) resources are far more difficult for a first-party administrator to remove --- they could, of course, remove the intermediate level 1 resource, but this may disrupt their own business activities. 
Table~\ref{tab:vulnerable_cat_perlevel} presents the proportion of websites that import at least one resource with a VTscore $\geq 10$. 
We separate resources into their level in the dependency chain. 
Interestingly, the majority of resources classified as suspicious are located at level 1 in the dependency chain (\ie they are explicitly trusted by the first-party). 
73\% of websites containing suspicious third-parties are ``infected'' via level 1. This suggests that these website operators are not entirely diligent in monitoring their third-party resources.  This might include websites that purposefully utilize such third-parties~\cite{ibosiola2018movie}.
Perhaps more important, the above leaves a significant minority of suspicious resources imported via \emph{implicit} trust (\ie level $\geq2$). In these cases, the first-party is potentially unaware of their presence. The most vulnerable category is news: over 15\% of news sites import \emph{implicitly} trusted resources from level 2 with a VTscore $\geq$ 10. Notably, among the 56 news websites importing suspicious JavaScript resources from trust level 3 and deeper, we find 52 loading advertisements from \texttt{adadvisor.net}. One possible reason is that ad-networks could be infected or victimized with malware to perform malvertising~\cite{malvertisingPaper, times_malvertising}.

\begin{table}[ht!]
\centering
\small
\tabcolsep=0.08cm
\scalebox{0.799} {
\begin{tabular}{l | c c  | c c | c c | c c | c c}
\toprule
& \multicolumn{2}{c}{{\bf All}} & \multicolumn{2}{c}{\bf News} & \multicolumn{2}{c}{\bf Sports} & \multicolumn{2}{c}{ \bf Entertainment} &  \multicolumn{2}{c}{\bf Forums}\\ \cline{2-11}
Lv.       & {All}     & {JS}     & All          & JS          & All      & JS    & All      & JS  & All      & JS  \\
\midrule
1	&	61.30\%	&	57.70\%	&	75.40\%	&	73.50\%	&	75.70\%	&	73.20\%	&	69.30\%	&	65.60\%	&	67.40\%	&	65.50\%	\\
2	&	5.20\%	&	2.20\%	&	13.40\%	&	5.60\%	&	11.10\%	&	3.70\%	&	8.60\%	&	4.10\%	&	9.10\%	&	4.05\%	\\
3	&	1.30\%	&	0.18\%	&	2.90\%	&	0.45\%	&	3.60\%	&	0.28\%	&	2.70\%	&	0.30\%	&	3.20\%	&	0.15\%	\\
4	&	0.22\%	&	$\leq$ 0.1\%	&	0.64\%	&	0.08\%	&	0.80\%	&	$\leq$ 0.1\%	&	0.70\%	&	0.08\%	&	0.60\%	&	0.00\%	\\
$\geq$ 5	&	$\leq$ 0.1\%	&	0	&	0.002	&	$\leq$ 0.1\%	&	0.001\%	&	$\leq$ 0.1\%	&	0.002\%	&	$\leq$ 0.1\%	&	$\leq$0.001\%	&	0.00\%	\\
\bottomrule
\end{tabular}
}
\caption{\small Proportion of top-200K websites importing resources classified as suspicious (with VTscore $\geq$ 10) at each level.} 
\label{tab:vulnerable_cat_perlevel}
\end{table}

Similar, albeit less extreme, observations can be made across Sports, Entertainment, and Forum websites. Briefly, Figure~\ref{fig:fp_resource_bar} also displays the categories of (suspicious) third-parties loaded at each level in the dependency chain --- it can be seen that the majority are classified as business. This is, again, because of several major providers classified as suspicious such as \texttt{convexity.net} and \texttt{charter.com}. 
Furthermore, it can be seen that the fraction of advertisement resources also increases with the number of levels due to the loading of further resources (\eg images).

\begin{figure}[ht!]
	\includegraphics[width=\columnwidth,height=4cm,keepaspectratio]{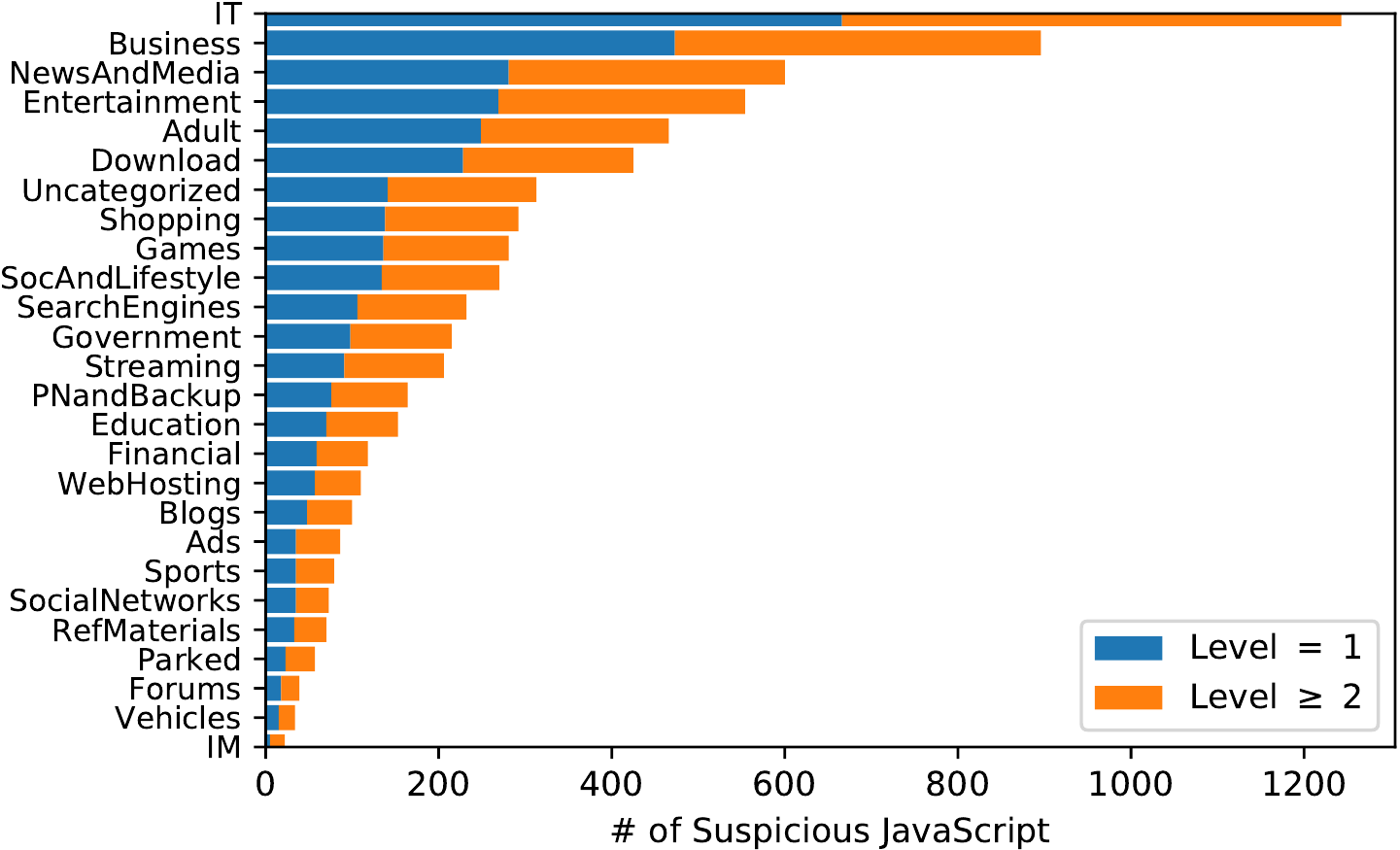}
	\caption{\small Breakdown of JavaScript resources based on category of domain.  Uncategorized category includes domain such as {\tt newmyvideolink.xyz} and {\tt cooster.ru}}
	\label{fig:js_dom_cat}
\end{figure}

Next, we again focus on JavaScript content as, when loaded, it can represent significant security risks: Our analysis is motivated by well known attack vectors underpinned by JavaScript, \eg malvertising \cite{malvertisingPaper}, malware injection and exploit kits redirection. These are exemplifed by the recent reporting that Equifax and TransUnion were hit by a third-party web analytics script \cite{Equifax_Malvertising, securityweekReport}.
Figure~\ref{fig:js_dom_cat} presents the breakdown of the domain categories hosting the JavaScript resources we observe. Clear trends can be seen, with IT (e.g., {\tt dynaquestpc.com}), Business (\eg {\tt vindale.com}),  News and Media (e.g., {\tt therealnews.com}), and Entertainment (\eg {\tt youwatchfilm.net}) dominating. 

\begin{figure}[t]
\centering

	\includegraphics[width=\columnwidth]{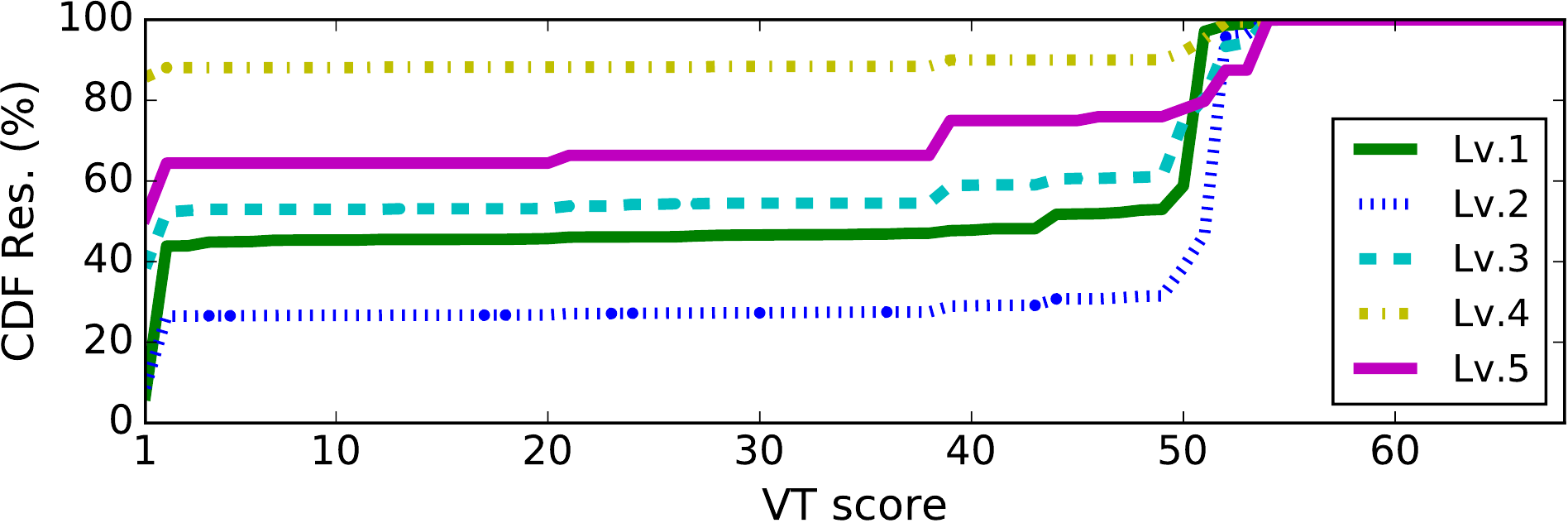}
	\caption{\small CDF of suspicious JavaScripts (VTscores $\geq10$) at different levels in the chain. }
	\label{fig:vt_cdf_level}
\end{figure}

Clearly, suspicious JavaScript resources cover a broad spectrum of activities. Interestingly, we observed that 70\% and 67\%, respectively, of Business (Web analytics) and Ads JavaScripts are loaded from level $\geq$ 2 in contrast to 17\% and 31\% of JavaScripts of Government and Shopping loaded at level 1.
We next strive to quantify the level of suspicion raised by each of these JavaScripts. Intuitively, those with higher VTscores represent a higher threat as defined by the 68 AV tools used by VirusTotal. Hence,  Figure~\ref{fig:vt_cdf_level} presents the cumulative distribution of the VTscores for all JavaScript resources loaded with VTscore $>0$. We separate the JavaScripts into their location in the dependency chain. Clear difference can be observed, with level 2 obtaining the highest VTscore (median 28). In fact, 78\% of the suspicious JavaScript resources loaded on trust level 2 have a VTscore $>52$ (indicating \emph{very} high confidence).

This is a critical observation since as mentioned earlier, while suspicious third-parties at level 1 can be ultimately removed by first-party website operators if flagged as suspicious, this is much more difficult for implicitly trusted resources further along the dependency chain. If the intermediate (non-suspicious) level 1 resource is vital for the webpage, it is likely that some operators would be unable or unwilling to perform this action. The lack of transparency and the inability to perform a vetting process on implicitly trusted loaded resources further complicates the issue. It is also worth noting that the VTscore for resources loaded further down the dependency chain is lower (\eg level 4). For example, 80\% of level 4 resources receive a VTscore below 5. This suggests that the activity of these resources is more contentious, with a smaller number of AV tools reaching consensus. 
It is impossible to state the reason for this, hence in Section~\ref{sec:banalysis}  we analyze the dynamic activities of these JavaScript content. 

\section{Analysis of Suspicious JavaScript resources}
\label{sec:banalysis}

JavaScript is arguably the riskiest resource to import as this has the potential to execute diverse functions (including the downloading of further resources). Thus, we next inspect the activities of the 7,166 JavaScript scripts that were classified as suspicious. 

\subsection{Methodology}
\begin{figure}[ht!]
\centering
	\includegraphics[width=\columnwidth]{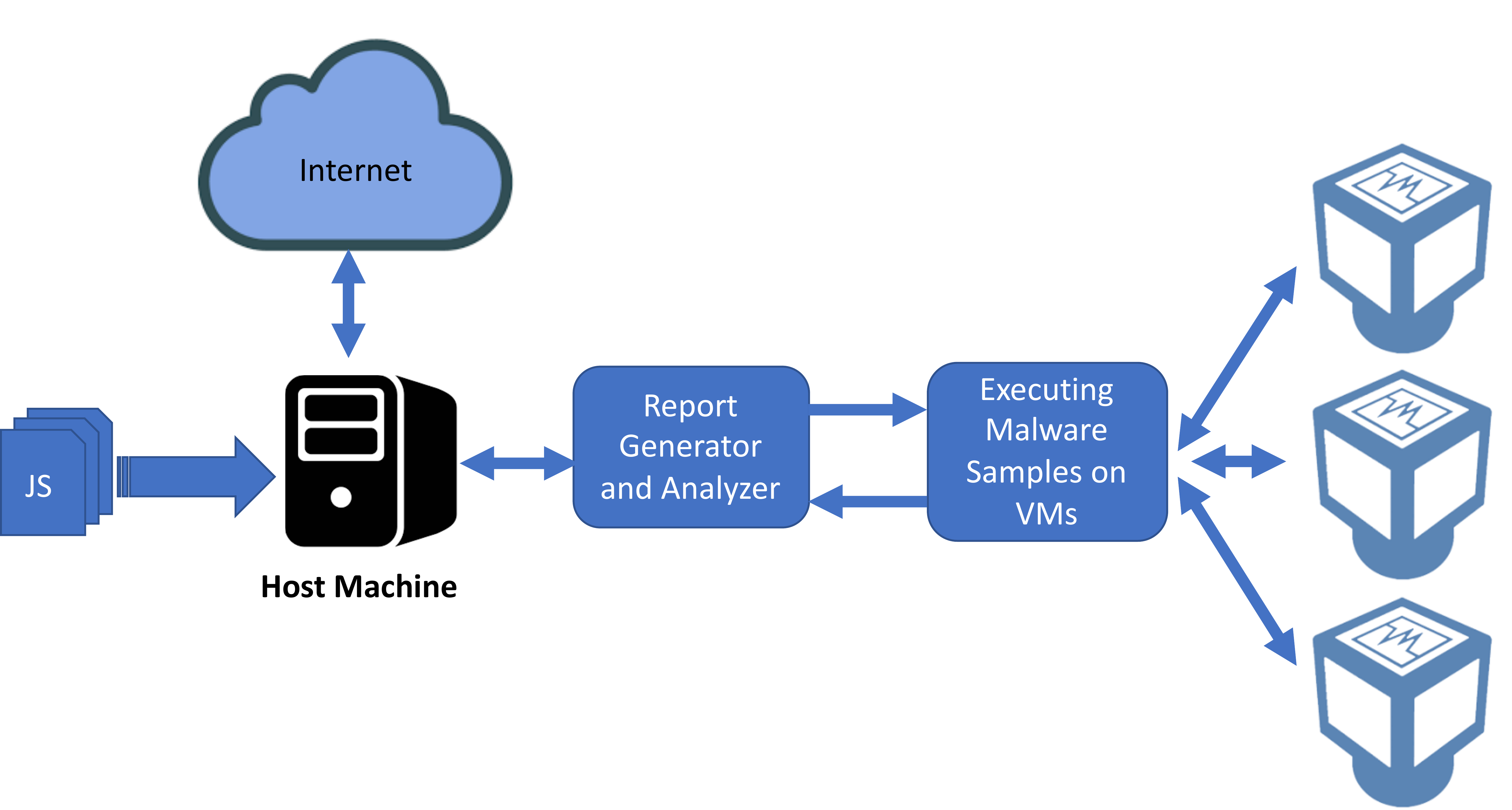}
	\caption{\small Overview of our testbed with a host machine running Cuckoo sandbox (v2.0.6).}
	\label{testbed}
\end{figure}

We use a dedicated testbed, depicted in Figure~\ref{testbed}, 
composed of three Virtual machines (VMs) that connect to the Internet via a computer that runs Cuckoo sandbox and \texttt{tcpdump}, respectively, to log all system-level events and to intercept all the traffic being transmitted between the virtual machines and the Internet. These VMs monitor already detected publishers using different browser user-agent configurations (IE 11 and Firefox 63.0.1) and cookie clearing strategies (``always clear cookies'').  This allows us to observe the traffic generated by each JavaScript code when it is rendered by the VMs' browsers. For instance, Cuckoo sandbox keeps track of network traffic generated; all types of file operations, memory changes, registry changes when the JavaScript code was loaded.  

For each test, first we create an HTML code and inject suspicious JavaScript code in $<script>$ tag and then load the HTML code in the browser of one of our VMs. Each test lasts 123.01$\pm$2.53 seconds. Prior to each test, we also ensure that the previous JavaScript code we experimented with is removed from the browser history as well as browser and VM's cache (\eg DNS cache) and we reboot the device to enforce the complete renewal of the VM. This analysis draws boundary for the suspicious behavior of JavaScript code during suspicious redirects and local file changes. We share the code and data for the testbed is at {\tt https://wot19submission.github.io}.  

\begin{table*}[ht!]
\centering
\small
\tabcolsep=0.04cm
\scalebox{0.9} {
\begin{tabular}{l c l l c r c c l}
\toprule
{\bf \#}	& {\bf Level} &	{\bf JavaScript}	& {\bf Category}&	{\bf VTscore}	&	{\bf A-Rank}	&	{\bf HTTP}	&	{\bf Domains}	&	{\bf Observed Behavior}	\\
\midrule
1	& Lvl-1 &	http://pinshan.com/js/union/new-play-1.js & Business	&	12	&	25,574	&	12	&	5	&	PUP activity, Installing Fake AV and mediaplayers	\\
2& Lvl-1	&	http://newyx.net/js/dui\_lian.js
 & Games	&	11	&	22,057 	&	12	&	6	&	Displaying annoying ads and perform click fraud	\\
3	& Lvl-1&	http://loxblog.com/fs/clocks/02.js
 & IT	&	10	&	86,505	&	10	&	3	&	Installing additional SW with elevated privileges	\\
4& Lvl-1	&	http://mecum.com/js/jquery.fancybox.pack.js
 & Business	&	13	&	51,897	&	9	&	3	& PUP activity, Installing Fake AV and mediaplayers	\\
5& Lvl-1	&	http://bubulai.com/js/xp.js
	& Enter. &	10	&	 117,261	&	9	&	5	& Displaying annoying ads and perform click fraud	\\
\midrule
1 &Lvl$\geq$2	&	http://yourjavascript.com/3439241227/blog.js & PNandBackup 	&	13	&	2,007,688	&	58	&	51	& 	Displaying annoying ads and perform click fraud	\\
2&Lvl$\geq$2	&	http://negimemo.net/alichina/login.js & SW Download	&	10	&	13,093,855	&	49	&	21	& Displaying annoying ads and perform click fraud		\\
3&Lvl$\geq$2	&	http://funday24.ru/js/c/funday-index.js	& News &	11	&	2,017,900	&	42	&	9	&	Displaying annoying ads and perform click fraud		\\
4&Lvl$\geq$2	&	http://netcheckcdn.xyz/optout/set/strtm.js & Business	&	14	&	18,064,762	&	42	&	7	&	Displaying annoying ads and perform click fraud	\\
5&Lvl$\geq$2	&	http://pushmoneyapp.com/js/main.js	 & Business &	17	&	8,757,970	&	41	&	6	& Installing additional SW with elevated privileges		\\

\bottomrule
\end{tabular}
}
\caption{\small Top 5 suspicious JSes, with at least 41 HTTP requests, found in the dependency chain (level 1 and levels$\geq$2) of 200K domains.}
\label{tab:http_req}
\end{table*}

\subsection{Results}

Our sandbox provides vantage onto the network activity generated by JavaScript resources, alongside any dropfile they generate. We next analyze the JavaScript along these two axes.

\subsubsection{Summary of Suspicious JavaScripts}

We begin by providing a brief overview of the most prominent JavaScript resources observed in our dataset. 44.7\% exist at level 1 (explicit trust) and 55.3\% at level $\geq$ 2 (implicit trust). 
Table~\ref{tab:http_req} provides a list of the JavaScript resources that generate the most HTTP requests. We separate them into implicit and explicitly trusted resources. 
The most typical purpose of these JavaScripts are downloading dropper files. Whereas those at level 1 tend to have lower VTscores, whereas VT classifies those at level $\geq$ 2 with a higher confidence. The most regularly observed JavaScript at level 1 is {\tt new-play-1.js}, a relatively highly ranked (22,574 Alexa) script which downloads dropfiles--executables such as Exploitkits, Trojans, Fake AV or mediaplayers performing suspicious activities. In contrast, at level $\geq$  2, it is {\tt blog.js}, which show annoying ads and involve in click fraud. 
Overall, we see that 99.5\% of JavaScript resources download dropfiles with 98.62\% of it involved in malvertising and click frauds. 

\subsubsection{Analyzing Network Activity }

\begin{figure*}[ht!]
\centering
	\subfloat[] 
	{\includegraphics[width=0.33\textwidth]{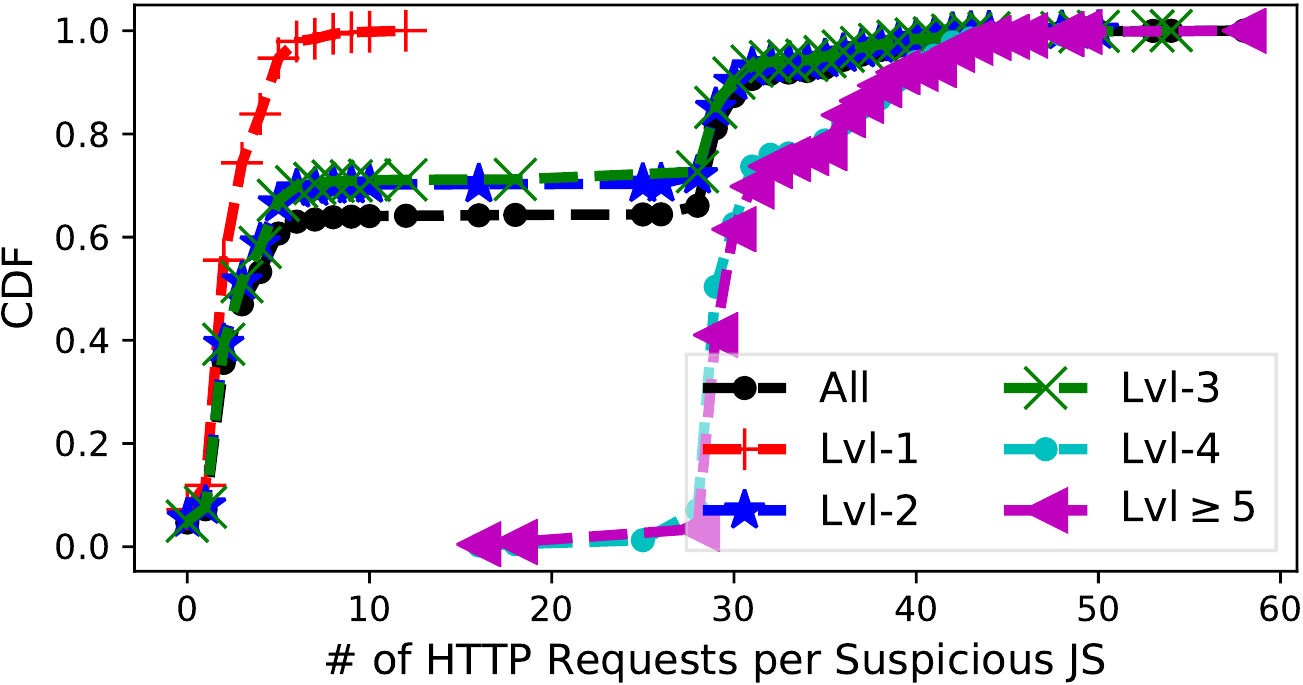}\label{subfig:httpreq_per_js}}
	\subfloat[]
	{\includegraphics[width=0.33\textwidth]{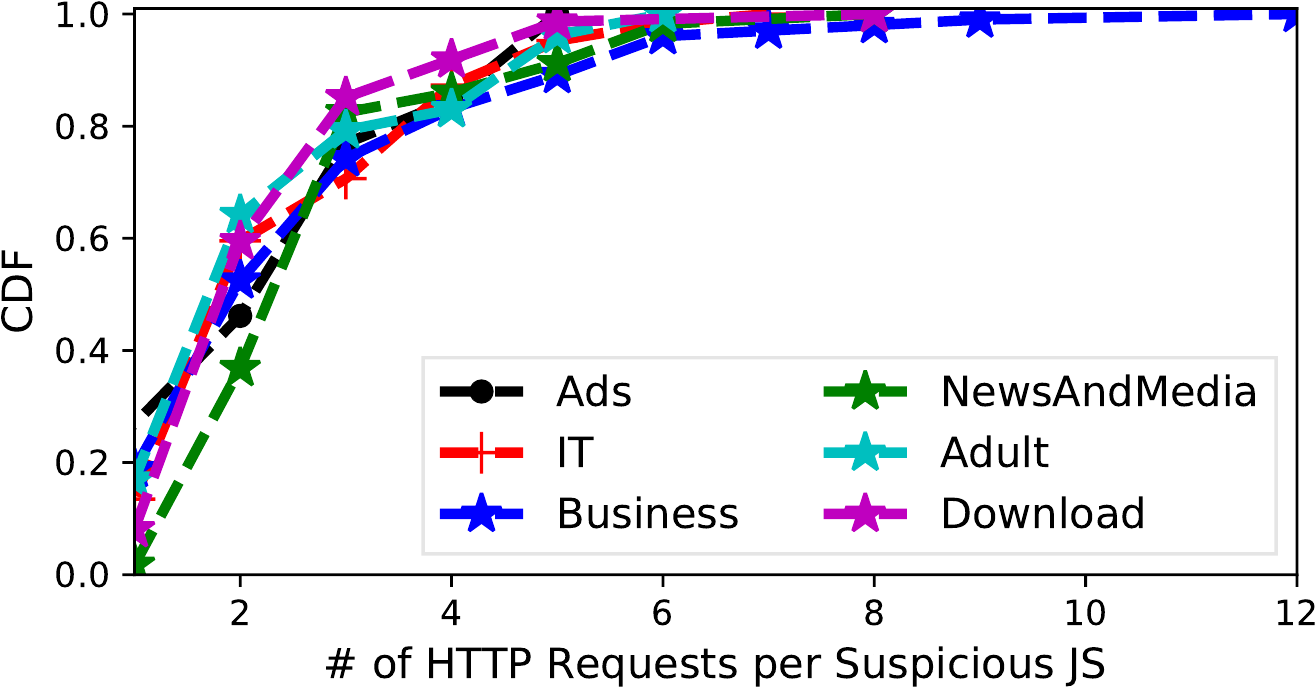}\label{subfig:http_req_by_js_lvl1}}
	\subfloat[] 
	{\includegraphics[width=0.33\textwidth]{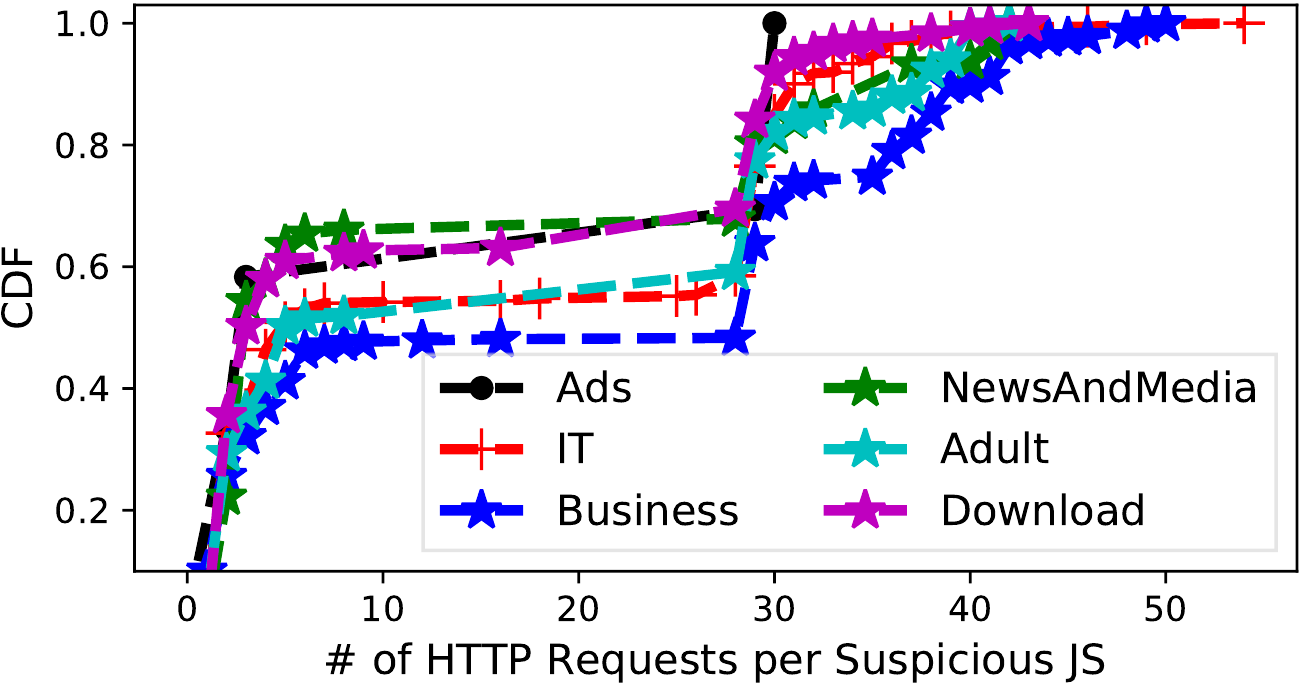}\label{subfig:http_req_by_js_lvl2}}
	\caption{\small CDFs of number of HTTP requests generated per suspicious JS viewed across categories of domains and dependency levels.}
		\label{fig:httpreq_per_js}
\end{figure*}

We observe significant network activity generated by the suspicious JavaScript resources within our testbed. 
Figure~\ref{fig:httpreq_per_js} presents the CDFs of the number of HTTP requests generated per suspicious JavaScript code. The figure splits the JavaScript into their respective positions in the dependency chains, and their categories (categories of third-party domains in which they originate from). Although 47\% of JavaScript resources generate fewer than 5 requests, there are notable differences among the different levels. JavaScript resources at level 1 generate the fewest HTTP requests (median 2), yet level $\geq$ 4 are extremely active (median 30). 36\% of the JavaScripts imported from level 5 generate at least 30 HTTP requests in contrast to 15\% of the JavaScripts sourced from level 2. 
This is in contrast to a typical behavior of legitimate JavaScripts that have been previously measured to generate on average 4 HTTP requests\cite{schneider2008new}. 
In essence, we observed that the further down the dependency chain, the more active the suspicious JavaScript. This is worrying as resources loaded further down the dependency chain are the most opaque to the website operator.

Figure~\ref{subfig:http_req_by_js_lvl1} and \ref{subfig:http_req_by_js_lvl2} also separate the JavaScript resources into their respective categories. Whereas those at level 1 (explicit trust) exhibit relatively similar traits across all categories, we find that those at level $\geq$ 2 (implicit trust) have far more divergence. Those classified as Business, IT or Adult are the most active, whereas News, Ads and Download generate the fewest. 
This is driven by the fact that most Business (\ie~ subcategory web-analytics) domains are continuously tracking dynamic content for a number of domains rather than, for example, JavaScripts codes from IT (\ie~ subcategory hosting providers), Adult, and Blogs domains, which source static resources for various domains~\cite{ikram2017towards}.
For instance, JavaScripts from Business category at level 1 generate on average 3 HTTP requests \vs 19 for those at level $\geq$ 2.

\begin{figure}[ht!]
 	\includegraphics[width=\columnwidth,height=4cm,keepaspectratio]{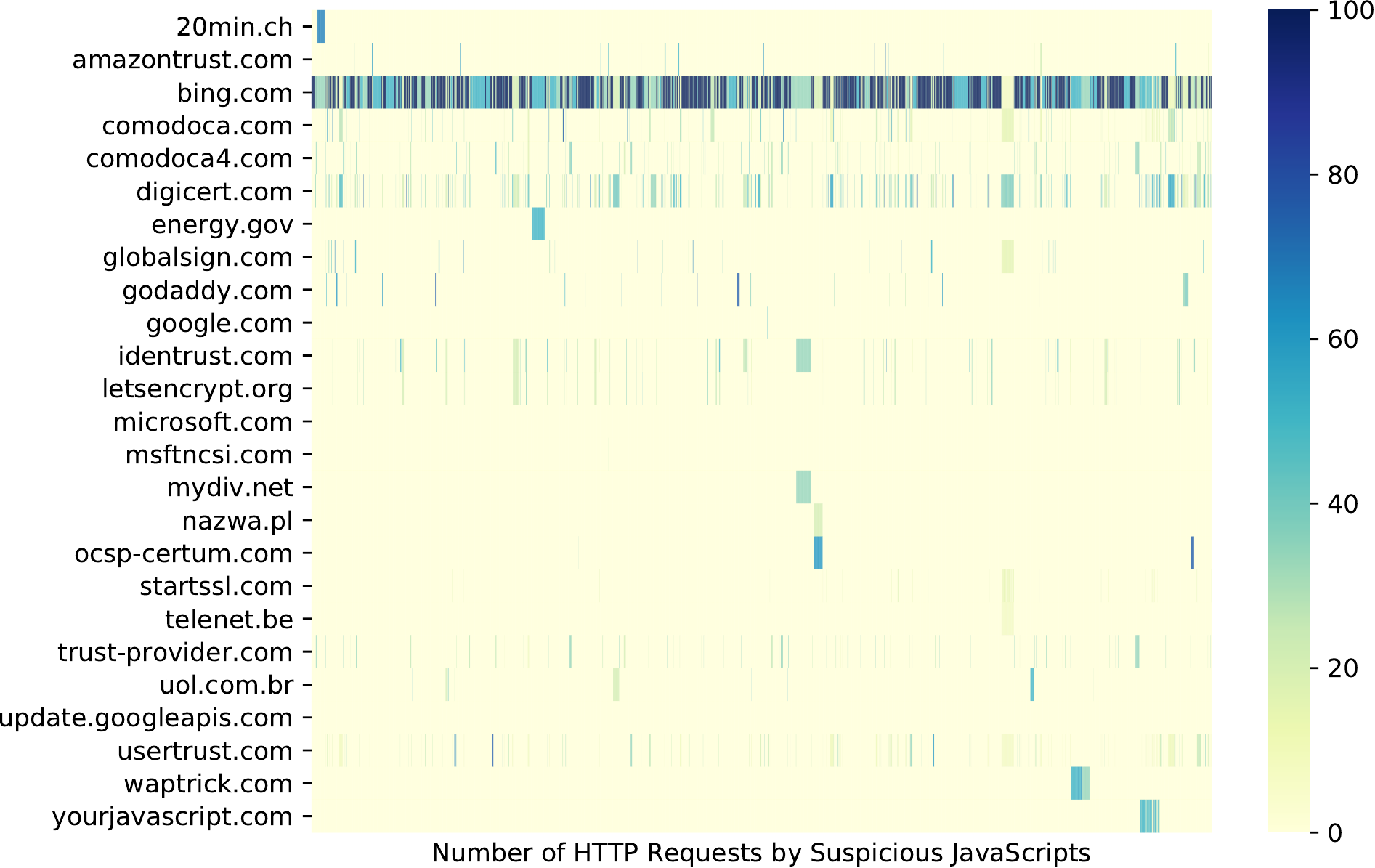}
	\caption{\small Heatmap of number of requests to domains by suspicious JavaScript codes and histogram of top 25 contacted domains by suspicious JavaScripts.}
	\label{fig:heatmap}
\end{figure}

A very important consideration here is to understand what are the domains that are targeted by these HTTP requests. For ease of presentation, we consider the top 25 domains targeted by the Suspicious JavaScript codes (in terms of total number of HTTP requests targeting them). 
Figure~\ref{fig:heatmap} presents a heatmap illustrating the number of requests to the top 25 domains by the suspicious JavaScript codes, with the the X-axis showing the suspicious JavaScript, the Y-axis listing the targeted domains and the heat defined by the fraction of requests that each JavaScript issued to each domain. We immediately see that most JavaScripts have a distinct preference towards a small set of domains. 
18\% of JavaScripts submit over 50\% of their HTTP requests to a single domain. One particularly popular domain is \texttt{bing.com}; 65\% of all JavaScripts access this domain at least once. A closer inspection of these JavaScript suggests that most of the ones targeting bing.com undertake some form of SEO poisoning (search poisoning) activities launching exploits to trick and elevate the ranking of some particular URLs in the results listings of the search engine~\cite{howard2010poisoned, invernizzi2012evilseed}.

\begin{figure} [ht!]
	\includegraphics[width=\columnwidth,height=4cm,keepaspectratio]{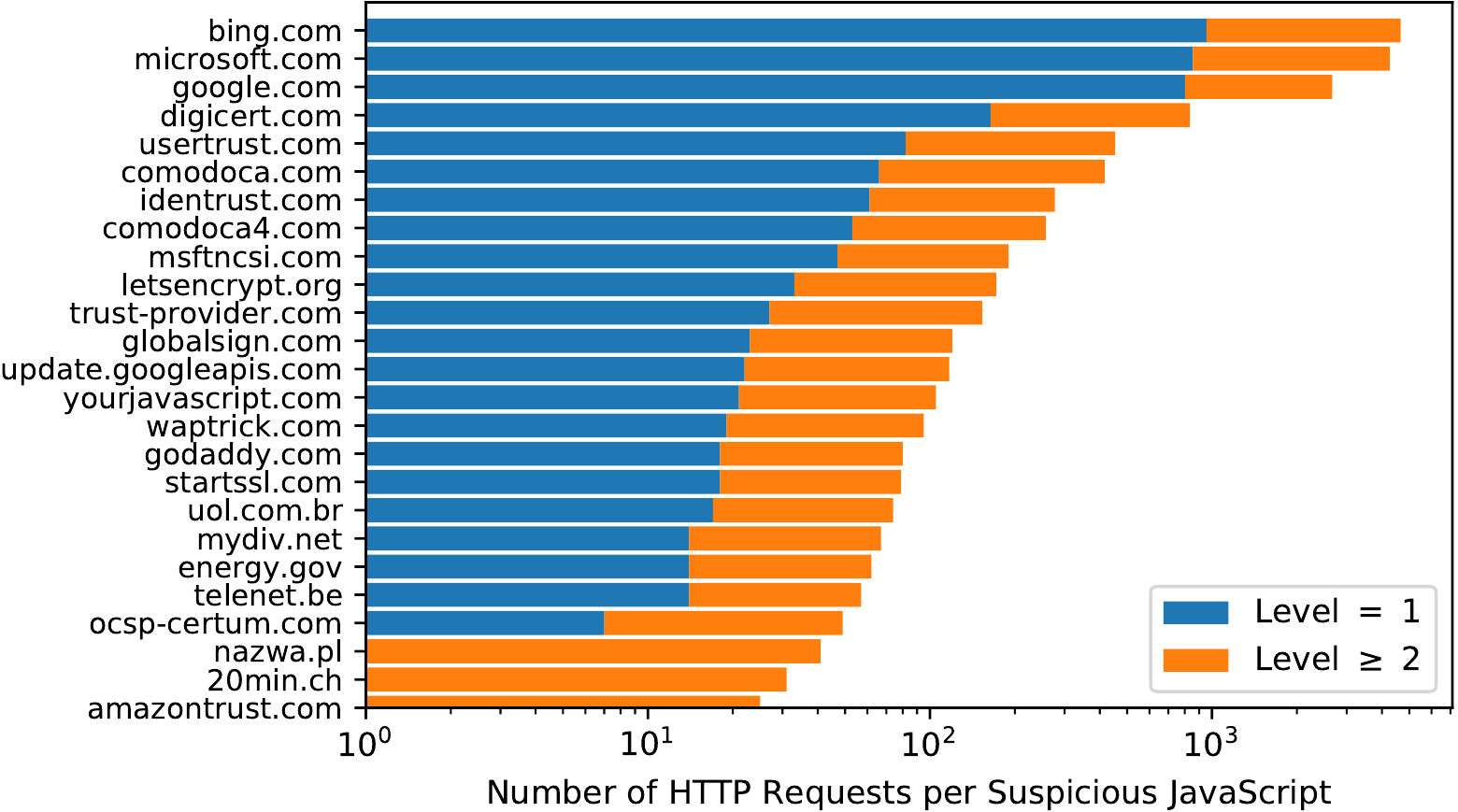}
	\caption{\small Number of HTTP requests launched to each domain.}
	\label{subfig:domain_count_barchart}
\end{figure}

Figure~\ref{subfig:domain_count_barchart} presents the count of HTTP requests across the top 25 targeted domains. We separate into JavaScripts at level 1 \vs level $\geq$ 2. Again, the figure shows that most commonly occurring HTTP fetch is to \texttt{bing.com}; In general, we observed that search engines are heavily targeted by scripting codes (to elevate certain URLs and websites as observed in previous research ~\cite{invernizzi2012evilseed}). Similar organizations are seen within this top ranking, \eg \texttt{microsoft.com}, \texttt{google.com}. We stress the importance of this observation as we recall that the JavaScript code is loaded on the first-party domain (either explicitly or implicitly). Hence the first-party domains might be involved in stealthy suspicious activities triggered by third-party domains the former are not even aware of. This observation is even more enlightened by our earlier observation that suspicious JavaScript are more active when at level $\geq$ 2 which is confirmed in Figure~\ref{subfig:domain_count_barchart}.         
We note the existence of various certification authorities and observed various JavaScript codes requesting encryption (SSL) methods and HTTPS requests that we did not investigate further. We leave this as future work.  

\begin{figure*}[ht!]
\centering
\subfloat[DropFiles by JSes]{\includegraphics[width=0.33\textwidth]{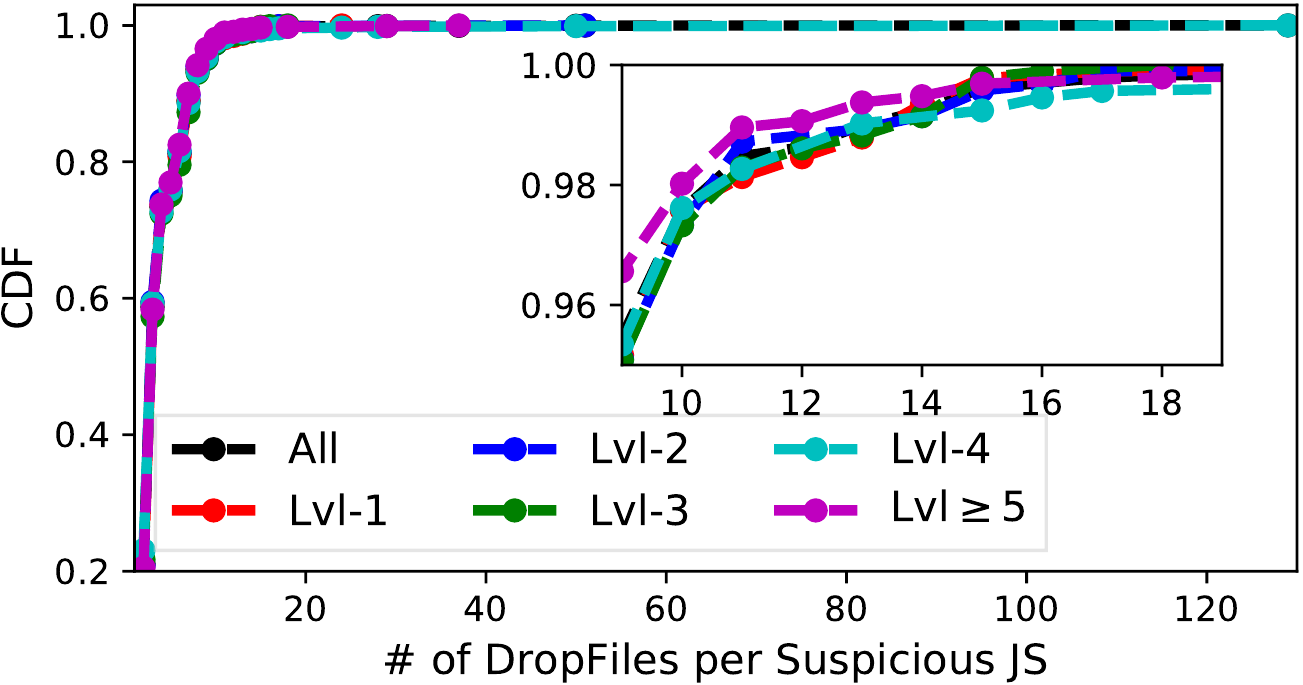}\label{subfig:dropfiles_per_js}}
\subfloat[DropFiles by JSes at Level = 1]{\includegraphics[width=0.33\textwidth]{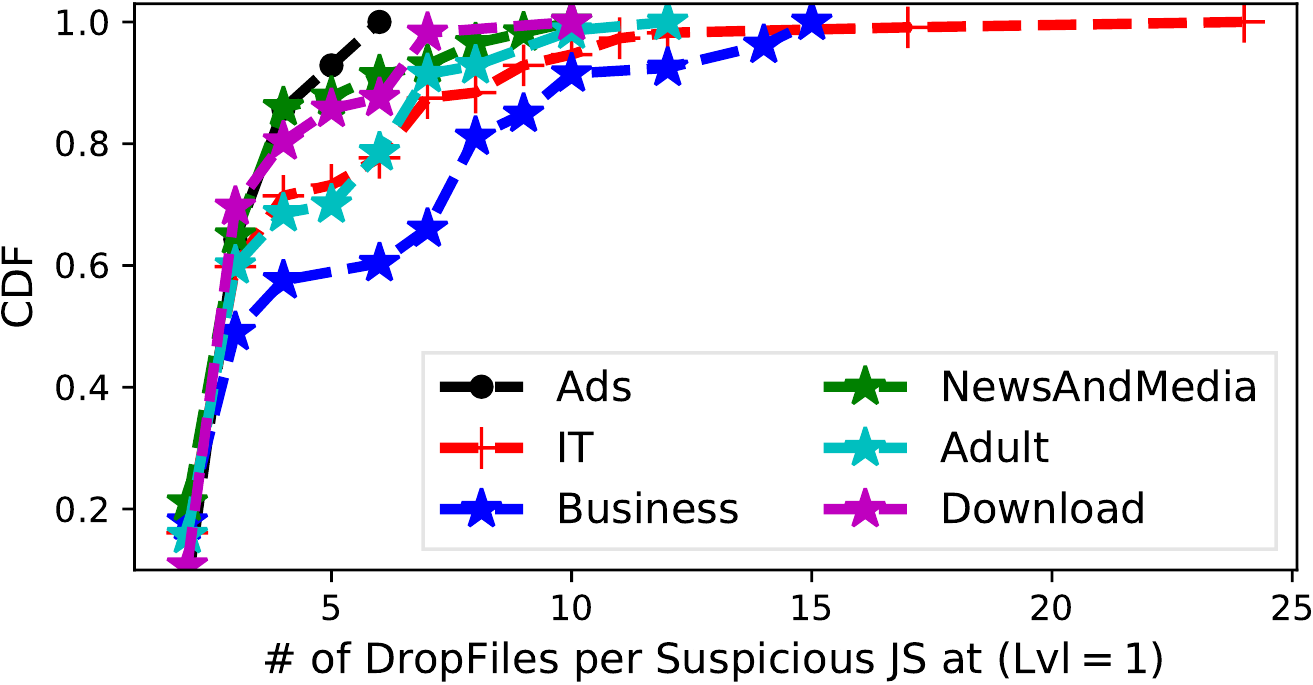}\label{subfig:dropfiles_per_js_lvl1}}
\subfloat[DropFiles by JSes at Level $\geq$ 2]{\includegraphics[width=0.33\textwidth]{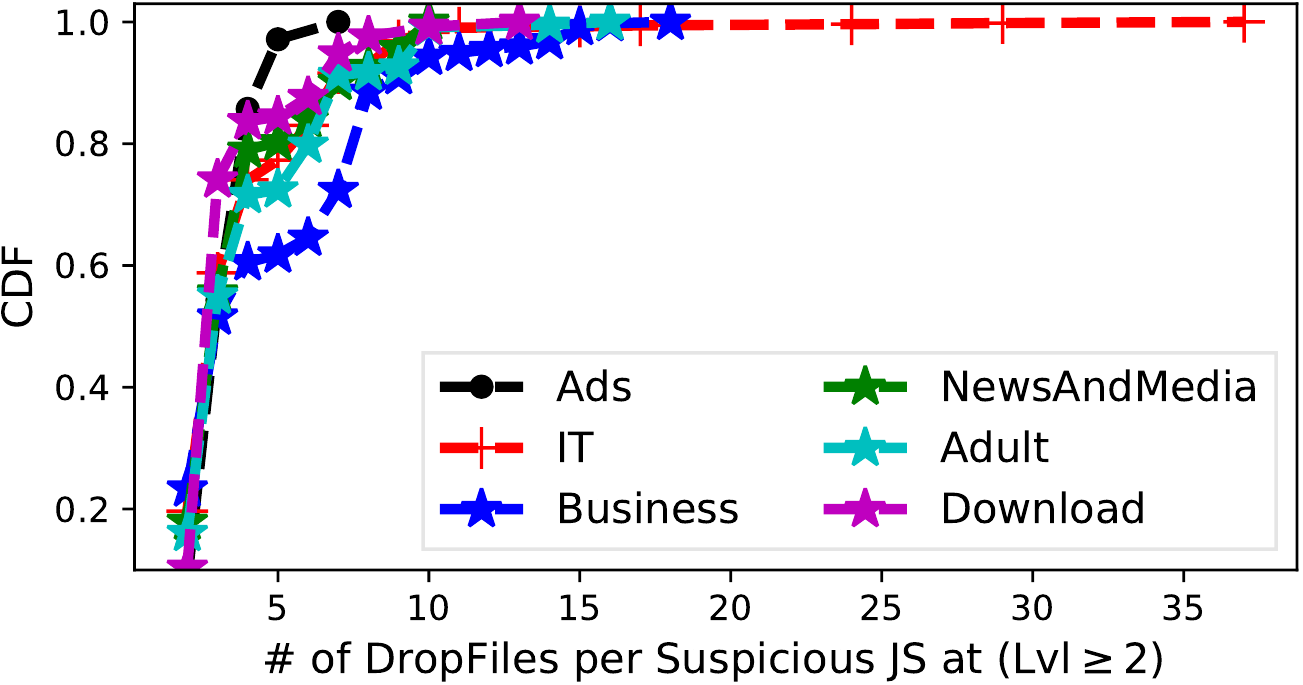}\label{subfig:dropfiles_per_js_lvl1}}
\caption{\small CDF of dropfiles downloaded and operated (i.e., read/write operation) by suspicious JavaScript codes.}
\label{fig:droppedfiles}
\end{figure*}

\subsubsection{Analyzing Dropfiles}
\label{subsubsec:adropfiles}

The above has revealed that a large number of suspicious JavaScript resources download files. The use of ``droppers'' is commonplace, and they are often used during the infection process~\cite{bencsath2012duqu}. For instance, these files can potentially contain the unpacked malware binary that could potentially present worrisome vulnerabilities. Hence, we next inspect the creation of local files by JavaScript. 

We find that 99.5\% of the analyzed JavaScript resources generate dropfiles, indicating that this is one of the most common used beavior of suspicious JavaScripts. Figure~\ref{fig:droppedfiles} depicts the distribution of the number of files dropped by the JavaScript content. We observe a significant number of active memory operations as depicted by the number of dropped files: remarkably, 22\% of JavaScript generate at least 8 files from memory. Table~\ref{tab:dropped_files} presents the top-10 most active, ranked according to the number of dropfiles, suspicious JavaScript.
Some of these JavaScript resources are extremely active. For example, {\tt xbigg.com/adv.js}, which is loaded at level 1, downloads 16 files. More worryingly, the propensity to download drop files actually increases further along the chain: the most active JavaScript, \url{http://yourjavascript.com/ 3439241227/blog.js} at level 4 downloads 129 files. 
It is also interesting to observe that the resources at level $\geq$ 2 also tend to have higher VTscores, indicating that their activities are blocked by a large number of virus checkers. 
The actual content of the files are quite diverse. 8\% are encrypted, and therefore cannot be examined. The remainder are 
Potentially Unwanted Programs (0.52\%), Exploitkits (0.36\%), Adware and Click Bots (98.62\%), and Trojan (0.50\%) according to VirusTotal reports. 
For instance, we observe that \url{videowood.tv/assets/js/poph.js} uses {\tt eval()}--JavaScript's dynamic loading method--to download and executes via {\tt 1832-fc204a9bcefeab3d.exe} (with VTscore=5)-- to take over web browser for displaying a wide range of annoying ads and to garner fraudulent clicks.

\begin{table*}[ht!]
\centering
\small
\tabcolsep=0.08cm
\scalebox{0.85} {
\begin{tabular}{l c l r c r  c l }
\toprule
{\bf $\#$ } & {\bf Level} & {\bf  JavaScript Code}   & {\bf  $\#$ of Drop files }  & {\bf $\%$ of Mal. DropFiles}  &{\bf  VTscore} & {\bf  Mal. Type}    & {\bf Observed Behavior }\\  
\midrule
1 &Lvl-1 & http://xbigg.com/adv.js & 16 & 64\% & {9}& AdCB &  Displaying annoying ads and perform click fraud\\
2 &Lvl-1& https://passback.free.fr/webmails/js/edito.js & 10 & 82\% & {4}& AdCB &  Displaying annoying ads and perform click fraud\\
3 &Lvl-1& http://via-midgard.info/engine/ajax/loginza.js & 10 & 89\% & {5}& AdCB &  Displaying annoying ads and perform click fraud\\
4 &Lvl-1& http://pokesnipe.de/js/app.min.js  & 10 & 90\% & {4}& Torjan & Installing additional SW with elevated privileges\\
5 &Lvl-1& http://hdvideo18.com/includes/ace.min.js&  8 &100\% & {3}& AdCB& Displaying annoying ads and perform click fraud\\
\midrule
1&Lvl$\geq$2 & http://yourjavascript.com/3439241227/blog.js & 129 & 20\% & {15}& AdCB &  Displaying annoying ads and perform click fraud\\
2&Lvl$\geq$2 & http://s2d6.com/js/globalpixel.js & 25 & 69\% & {11}& AdCB & Displaying annoying ads and perform click fraud\\
3&Lvl$\geq$2 & http://cmsdude.org/wp-includes/js/jquery/jquery.js & 12& 71\% & {11}& AdCB &  Displaying annoying ads and perform click fraud\\
4 &Lvl$\geq$2& http://widih.com/js/like.js & 10 &90\%& {10}& PUP &  PUP activity, Installing Fake AV and mediaplayers\\
5&Lvl$\geq$2 & http://pichak.net/blogcod/clock/04/clock.js & 8 & 100\% & {10} & Exploitkit & Installing Exploitkit and performing Web redirects\\
\bottomrule
\end{tabular}
}
\caption{\small Top 5 suspicious JavaScript codes with dropped files at explicit and implicit dependency level. AdCB means Adware and Click Bots. }
\label{tab:dropped_files}
\end{table*}

\section{Related Work}
\label{sec:rwork}
There has been a wealth of research into the utilisation and exploitation of third-parties~\cite{ikram2017towards} \cite{Nikiforakis2012} \cite{Lauinger2017} \cite{Ikram:2014:IJW} \cite{SU201983} \cite{ipccc2018}. 
Falahrastegar \etal~\cite{falahrastegar2014anatomy} inspected the use of third-parties across top Alexa websites, exploring how third-party operators differ based on region. 
Nikiforakis \etal \cite{Nikiforakis2012} demonstrated in 2012 that large proportions of websites rely on JavaScript libraries hosted on ill-maintained external web servers, making JavaScript exploits trivial. Lauinger \etal \cite{Lauinger2017} led a further study, classifying sensitive libraries and the vulnerabilities caused by them. Gomer \etal \cite{Gomer2013} analysed users' exposure to tracking in the context of search, showing that 99.5\% of users are tracked by popular trackers within 30 clicks. Further, Hozinger \etal \cite{Holzinger:ccs2016} found 61 JavaScript exploits and empirically  defined three main attack vectors. 

Our work differs quite substantially from these in that we are not interested in the JavaScript code itself, nor the simple presence of third-party domains in a webpage. Instead, we are interested in \emph{how} third-parties are loaded, and their use of ``implicit'' trust (\ie dependency chains). 
In contrast to our work, these prior studies ignore the presence of dependency chains and treat all third-parties as ``equal'', regardless of where they are loaded in the dependency chain. 
Closer to our own work is Bashir \etal \cite{Bashir2016}, who studied websites' resource inclusion trees and analyzed retargeted ads using crowdsourcing. This allowed them to identify and classify ad domains, as well as predominant cookie matching partners in the ad exchange environment. Our study is far broader, and sheds light on dependency chains across many different types of websites rather than simply inspecting advertisements. More related is Kumar \etal \cite{Kumar2017}, who recently characterized websites' resource dependencies on third-party services. In-line with our work, they found that dependency chains are widespread. This means, for example, that 55\% of websites are prevented from fully migrating to HTTPS by their dependencies. Their focus was not, however, on identifying suspicious or malicious activities. To the best of our knowledge, this paper is the first study to analyze the role of implicit trust from a security perspective to better understand the role of dependency chains in loading suspicious third-party content.

\section{Concluding Remarks}
\label{sec:conclusion}

This paper has explored dependency chains in the web ecosystem. Inspired by the lack of prior work focusing on how resources are loaded, we found that over 40\% of websites \emph{do} rely on implicit trust. 
Although the majority (84.91\%) of websites have short chains (with levels of dependencies below 3), we found first-party websites with chains exceeding 30 levels. Of course, the most commonly \emph{implicitly} trusted third-parties are well known operators (\eg \texttt{doubleclick.net}), but we also observed various less known implicit third-parties.
We hypothesised that this might create notable attack surfaces. 
To confirm this, we classified the third-parties using VirusTotal to find that 1.2\% of third-parties are classified as potentially malicious. Worryingly, our confidence in the classification actually increases for implicitly trusted resources (\ie trust level2), where 78\% of JavaScripts have a VTscore $>52$. 
These resources have remarkable reach --- largely driven by the presence of highly central third-parties, \eg \texttt{google-analytics.com}.
With this in mind, we performed sandboxes experiments on the suspicious JavaScript to understand their actions. We witnessed extensive download activities, much of which consisted of dropper files and malware, which was being installed on the machine. It was particularly worrying to see that JavaScript resources loaded at  level $\geq$ 2 in the dependency chain tended to have more aggressive properties, particularly as exhibited by their higher VTscore. 
This exposes the need to tighten the loose control over indirect resource loading and implicit trust: it creates exposure to risks such as malware distribution, SEO poisoning, malvertising and exploit kits redirection. 
We argue that ameliorating this can only be achieved through transparency mechanisms that allow web developers to better understand the resources on their webpages (and the related risks).

This is only the first step in our research agenda. Most notably, we wish to perform longitudinal measurements to understand how these metrics of maliciousness evolve over time. We are particularly interested in understanding the (potentially) ephemeral nature of threats. Without this, we are reticent to draw long-term conclusions. Another line of work is understanding how level $\geq$ 1 JavaScript content creates inter-dependencies between websites. This is particularly noteworthy among hypergiants (\eg Google), who are present on a large number of first-party websites. We intend to perform graph analysis to understand how removing these hypergiants may impact the presence of interconnected suspicious third-parties. Finally, by opening our collected datasets and experiments code and scripts to the wider research community, we hope this paper has shed some light on an important security consideration of today's Web and we call for further collaboration and research to help address this.

\balance
\bibliographystyle{abbrv}
\bibliography{biblo}

\end{document}